\documentclass[]{aa}  

\usepackage[utf8]{inputenc}
\usepackage{bm}
\usepackage{algpseudocode}
\usepackage{algorithm}
\usepackage{amsmath}
\usepackage{xcolor}
\usepackage{subfig}
\usepackage{comment}
\DeclareMathOperator*{\argmax}{arg\,max}

\newcommand{\action}{\boldsymbol{a}_t}
\newcommand{\obs}{\boldsymbol{o}_t}
\newcommand{\st}{\boldsymbol{s}_t}
\newcommand{\stp}{\boldsymbol{s}_{t+1}}

\usepackage{graphicx}
\usepackage{txfonts}
\begin{document} 

   \title{Towards on-sky adaptive optics control using reinforcement learning}

   \subtitle{Model-based policy optimization for Adaptive Optics}

   \author{J. Nousiainen
          \inst{1,3}
          \and
          C. Rajani\inst{2}
          \and M. Kasper\inst{3}
          \and T. Helin\inst{1}
          \and
          S.~Y.~Haffert\inst{4,5}
   \and 
   C. Vérinaud \inst{3}
   \and 
   J.~R.~Males\inst{4}
   \and 
   K.~Van Gorkom\inst{4}
   \and
   L.~M.~Close\inst{4}
   \and
   J. D. Long\inst{4}
   \and
   A. D. Hedglen\inst{4,6}
   \and
   O. Guyon\inst{4,6,7,8}
   \and
   L. Schatz\inst{9}
   \and
   M. Kautz\inst{4,6}
   \and
   J. Lumbres\inst{4,6}
   \and 
   A. Rodack\inst{4,6}
   \and
   J. M. Knight\inst{4,6}
   \and
   K. Miller\inst{9}
          }

   \institute{Lappeenranta--Lahti University of Technology, \email{jalo.nousiainen@lut.fi}
         \and
             University of Helsinki, Department of Computer Science \and European Southern Observatory
        \and
        University of Arizona, Steward Observatory, Tucson, Arizona, United States
        \and
        NASA Hubble Fellow
        \and
        Wyant College of Optical Science, University of Arizona, 1630 E University Blvd, Tucson, AZ 85719, USA
        \and
        Astrobiology Center, National Institutes of Natural Sciences, 2-21-1 Osawa, Mitaka, Tokyo, JAPAN
        \and
        National Astronomical Observatory of Japan, Subaru Telescope, National Institutes of Natural Sciences, Hilo, HI 96720, USA
        \and
        Kirtland Air Force Base, Air Force Research Laboratory, Albuquerque, NM, USA
             }

   \date{Received xx xx, 2022; accepted xx xx, 2022}

 
  \abstract
   {The direct imaging of potentially habitable Exoplanets is one prime science case for the next generation of high contrast imaging instruments on ground-based extremely large telescopes. To reach this demanding science goal, the instruments are equipped with eXtreme Adaptive Optics (XAO) systems which will control thousands of actuators at a framerate of kilohertz to several kilohertz. Most of the habitable exoplanets are located at small angular separations from their host stars, where the current control laws of XAO systems leave strong residuals.}
   {Current AO control strategies like static matrix-based wavefront reconstruction and integrator control suffer from temporal delay error and are sensitive to mis-registration, i.e., to dynamic variations of the control system geometry. We aim to produce control methods that cope with these limitations, provide a significantly improved AO correction, and, therefore, reduce the residual flux in the coronagraphic point spread function (PSF).}
   {We extend previous work in Reinforcement Learning (RL) for AO. The improved method, called PO4AO, learns a dynamics model and optimizes a control neural network, called a policy. We introduce the method and study it through numerical simulations of XAO with Pyramid wavefront sensing for the 8-m and 40-m telescope aperture cases. We further implemented PO4AO and carried out experiments in a laboratory environment using Magellan Adaptive Optics eXtreme system (MagAO-X) at the Steward laboratory.}
   {PO4AO provides the desired performance by improving the coronagraphic contrast in numerical simulations by factors 3-5 within the control region of DM and Pyramid WFS, both in simulation and in the laboratory. The presented method is also quick to train, i.e., on timescales of typically 5-10 seconds, and the inference time is sufficiently small (< ms) to be used in real-time control for XAO with currently available hardware even for extremely large telescopes.}
   {}


   \maketitle
%
\section{Introduction}
The study of extrasolar planets (exoplanets) and exoplanetary systems is one of the most rapidly developing fields of modern astrophysics. More than 3000 confirmed exoplanets have been identified mainly through indirect methods by NASA’s Kepler mission \footnote{Exoplanet Orbit Database: http://exoplanets.org/}. High-contrast imaging (HCI) detections are mostly limited to about a dozen very young and luminous giant exoplanets (e.g., \citealp{2010Natur.468.1080M,2009A&A...493L..21L,2015Sci...350...64M}) due to the challenging contrast requirements at a fraction of an arcsecond angular distance from the star which could be a billion times brighter than the exoplanet.

HCI aims at optically separating exoplanet light from stellar light, thereby dramatically increasing the signal-to-noise ratio (S/N) over the one provided by indirect methods. However, significant advances in HCI technology are needed to address two major scientific questions, the architectures of outer planetary systems, which remain essentially unexplored (e.g., \citealp{2015ApJ...807...45D}, \citealp{2019ApJ...874...81F}), and the atmospheric composition of small exoplanets outside the solar system which is especially interesting because it addresses the question of habitability and life in the universe.

For ground-based observations, HCI combines eXtreme Adaptive Optics (XAO, e.g. \citealp{guyon2005limits, guyon2018extreme}) and coronagraphy \citep{2012SPIE.8442E..04M} with a way to distinguish stellar quasi-static speckles (QSS) produced by imperfect instrument optics from the exoplanet such as spectral- and angular differential imaging \citep{2004ApJ...615L..61M, 2006ApJ...641..556M} or high-dispersion spectroscopy \citep{2015A&A...576A..59S}. With an optimized instrument design, the XAO residual halo may be the dominant source of noise \citep{2021A&A...646A.150O}. Therefore, minimizing the XAO residuals is a key objective for ground-based HCI. 

AO systems typically run in a closed-loop configuration, where the WFS measures the wavefront distortions after DM correction. The objective of this control loop is to minimize the distortions in the measured wavefront, i.e., the residual wavefront, which, in theory, corresponds to minimizing the speckle intensity in the post-coronagraphic image. In the case of a widely used integrator controller, temporal delay error and photon noise usually dominate the wavefront error budget in the spatial frequency regime controlled by the DM \citep{guyon2005limits, fusco2006high}. A big part of the turbulence is presumably in frozen flow considering the millisecond time scale of AO control, and hence a significant fraction of wavefront disturbances can be predicted \citep{poyneer2009experimental}. Therefore, control methods that use past telemetry data have shown a significant potential for reducing the temporal error and photon noise \citep{males2018ground, guyon2017adaptive, correia2020performance}. Further, real systems suffer from dynamic modeling errors such as misregistration \citep{heritier2018new}, optical gain effect for the Pyramid WFS \citep{korkiakoski2008improving, deo2019telescope}, and temporal jitter \citep{poyneer2008predictive}. Combined, these errors lead to a need for external tuning and re-calibration of a standard pseudo-open-loop predictive controller to ensure robustness.

\begin{figure}[h]
    \centering
    \includegraphics[width=7.7cm]{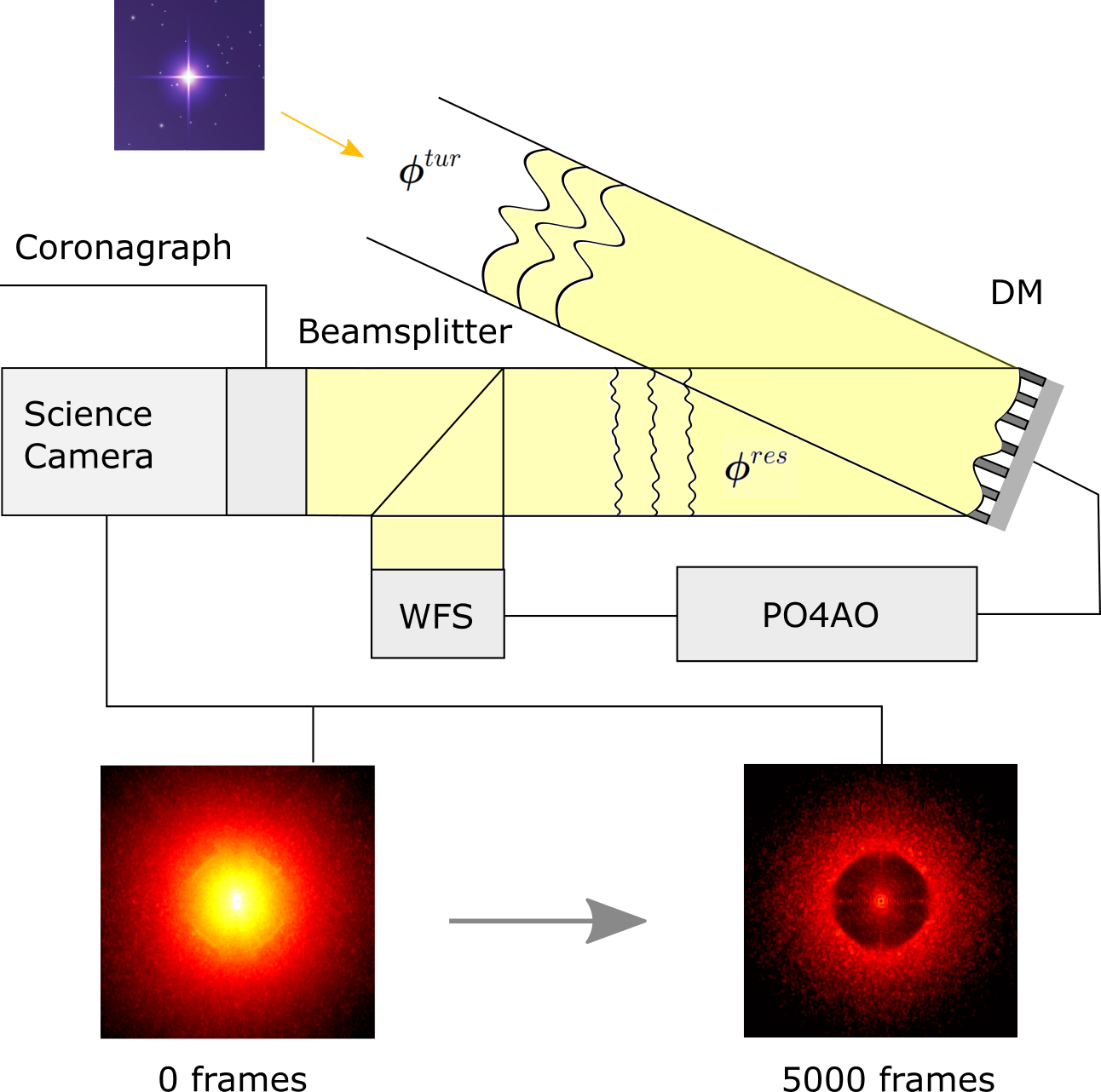}
    \caption{Overview of the AO control loop and the performance of the PO4AO. The method, PO4AO, feeds actions to the environment, observes the outcome, and then improves the control regarding the reward. Starting from a random behavior at first (frame 0), the method learns a predictive control strategy in only 5000 frames of interaction. }
    \label{fig:overview}
\end{figure}

An up-and-coming field of research aimed at improving AO control methods is the application of fully data-driven control methods, where the control voltages are separately added to the learned control model \citep{nousiainen2021adaptive, landman2020self, landman2021self, haffert2021data, haffert2021data1, Pou:22}. A significant benefit of fully data-driven control in closed-loop is that it does not require an estimate of the system's open-loop temporal evolution and that it is, therefore, insensitive to pseudo-open-loop reconstruction errors, such as the optical gain effect \citep{haffert2021data}. In particular, reinforcement learning (RL) has also been shown to cope with temporal and misregistration errors \citep{nousiainen2021adaptive}. RL is an active branch of machine learning that learns a control task via interaction with the environment. The principal idea is to let the method feed actions to the environment, observe the outcome, and then improve the control strategy regarding the long-term reward. The reward is a predefined function giving a concrete measure of the method's performance. By learning this way, RL methods do not require accurate models of the components in the control loop and, hence, can be viewed as an automated approach for control.

Previous work in RL-based adaptive optics control has focused on either controlling DM modes using model-free methods that learn a \emph{policy} $\pi_\theta : s_t \mapsto a_t$ parameterized by $\theta$ that maps observations/states $s_t$ into actions $a_t$ directly \citep{landman2020self, landman2021self, Pou:22}, or using model-based methods that employ a planning step to compute actions \citep{nousiainen2021adaptive}. The model-free methods have the advantage of being fast to evaluate, as the learned policies are often neural networks that support sub-millisecond inference. However, they suffer from the large space of actions resulting from the number of actuators that need to be controlled in adaptive optics systems -- learning to control each actuator simultaneously with a model-free method is difficult. On the other hand, model-based RL approaches benefit from being simple to train using even off-policy data, i.e., data obtained, while using a different (e.g., classical integrator) control method. A Model-based method may only need hundreds of iterations while a model-free algorithm such as the Policy Gradient method may need millions of iterations \citep{janner2019trust}. However, the planning step of model-based RL is often iterative and could, therefore, be too slow for AO control, even with expensive hardware \citep{nousiainen2021adaptive}.

In this paper, we unify the approaches described above by learning a dynamics model and using the model to train a policy that is fast to evaluate and scales to control all actuators in a system. We call this hybrid algorithm Policy Optimization for Adaptive Optics (PO4AO). We do this by employing an end-to-end convolutional architecture for the policy, leveraging the differentiable nature of the chosen reward function, and directly backpropagating through trajectories sampled from the model. Our method scales to sub-millisecond inference, and we present promising results in both a large pyramid-sensor-based simulation and a laboratory setup using MagAO-X \citep{males2018magao}, where our method is trained from scratch using interaction.

\section{Related work}

The AO control problem differs from the typical control problems considered by modern RL research. The main challenges of AO control are two-fold: first, the control space is substantially larger than in classical RL literature and is typically parameterized by 500 to 10000 degrees of freedom (DoF). Secondly, the state of the system is observed through an \emph{indirect measurement}, where the related inverse problem is not well-posed. On the bright side, it has been observed in the literature that simple differentiable reward functions with a relatively short time horizon can lead to good performance \citep{nousiainen2021adaptive}. 

Recently, progress has been made towards full reinforcement learning-based adaptive optics control. \cite{landman2020self} use the model-free recurrent deterministic policy gradient algorithm to control the tip and tilt modes of a DM and a variation of the method to control a high order mirror in the special case of ideal wavefront sensing. \cite{Pou:22} implemented a model-free multi-agent approach to control a 40 x 40 Shack-Harmann-based AO system and analyzed the robustness against noise and variable atmospheric conditions. On the other hand, \cite{nousiainen2021adaptive} present a model-based solution that learns a dynamics model of the environment and uses it with a planning algorithm to decide the control voltages at each timestep. This method shows good performance but requires heavy computation at each control loop iteration, which will be a problem in future generations of instruments with more actuators per DM. PO4AO aims for the best of both worlds: it requires only a small amount of training data and has a high inference speed, capable of scaling to modern telescopes. Further, we analyze the performance of our method in different noise levels and varied wind conditions combined with non-linear wavefront sensing.  

In RL terms, model-based policy optimization is an active area of research. Work that tackles the full reinforcement learning problem without assuming a known reward function includes \cite{heess2015learning}, and \cite{janner2019trust}. In contrast, PILCO and the subsequent deep PILCO \citep{deisenroth2011pilco, gal2016improving} are methods that directly backpropagate through rewards. Our method is similar to deep PILCO in the sense that it learns a neural network policy from trajectories sampled from a neural network dynamics model. 

In addition, significant progress has also been made in AO control methods outside RL and fully data-driven algorithms. Linear-quadratic-Gaussian control (LQG) based methods have been studied in \cite{kulcsar2006optimal, paschall1993linear, gray2012ensemble, conan1a2011integral, correia2010adapting,correia2010optimal, correia2017modeling}, sometimes combined with machine learning for system identification \citep{sinquin2020sky}. Predictive controllers have been studied in \cite{guyon2017adaptive, poyneer2007fourier, dessenne1998optimization,van2017performance, van2019impact}. Methods vary from linear filters to filters operating on single modes (such as Fourier modes) to neural network approaches \citep{swanson2018wavefront, sun2017bayesian, liu2019using, wong2021predictive}. Predictive control methods have also been studied in a closed-loop configuration. \cite{males2018ground} address a closed-loop predictive control's impact on the postcoronagraphic contrast with a semianalytic framework. \cite{swanson2021closed} studied closed-loop predictive control with NNs via supervised learning, where a NN is learned to compensate for the temporal error.

Finally, other RL-based methods have been developed for different types of AO. In order to mitigate alignment errors in calibration, a deep-learning control model was proposed in \cite{xu2019deep}. A model-free RL method for wavefront sensorless AO was studied in \cite{ke2019self}. The method is shown to provide faster corrections speed than a baseline method assuming a relatively low-order AO system, while our work focuses on the case of XAO for HCI.



\section{RL applied to AO}
\label{sec:prelim}

Since we introduce a novel approach (RL) to the field of AO, we present hereafter some of the standard notations and terms used in RL. The de facto mathematical framework for modeling sequential decision problems in the field of RL is the Markov Decision Process (MDP). An MDP is a discrete-time stochastic process which, at time step $t$, is in a state $\st \in \mathcal{S}$ where $\mathcal{S}$ is the set of all possible states. The decision-maker then takes an \emph{action} $\action \in \mathcal{A}$ (again, $\mathcal{A}$ is the set of possible actions) based on the current state, and the \emph{environment} changes to the next state $\stp$. As the transition dynamics $(\action, \st) \mapsto \stp$ is random in nature (influenced e.g. by the turbulence evolution) it is represented here by the conditional probability density function $p(\stp | \st, \action)$ \footnote{The initial state is drawn from the initial state distribution $\boldsymbol{s}_0 \sim p_0(\boldsymbol{s}_0)$}. At each timestep a \emph{reward} $R_t = r(\st, \action)$ is also observed, which is a (possibly stochastic) function of the current state and action. The modeler usually designs the reward to make the decision-maker produce some favorable behavior (e.g., correcting for turbulence distortions).

The actions our decision-maker takes are determined by a \emph{policy} $\pi_\theta : \st \mapsto \action$, which is a function that maps states into actions. For example, the matrix-vector multiplier (MVM) can be viewed as a policy, taking a wavefront sensor measurement as input and outputting the control voltages. The objective of reinforcement learning is to find a policy such that
\begin{align}
    \label{eq:obj_of_rl}
    \argmax_\theta \mathbb{E}_{p_\theta({\bf s}_0,..., {\bf s}_T)} \left[ \ \sum_{t=0}^T r(\bm s_t,\pi_\theta(\bm s_{t})) \ \right],
\end{align}
where 
\begin{equation*}
p_\theta({\bf s}_0,..., {\bf s}_T) = p_0(\bm s_0) \prod_{t=1}^T p(\st | {\bf s}_{t-1}, \pi_\theta({\bf s}_{t-1}))
\end{equation*}
with the initial distribution $\bm s_0 \sim p_0$ and convention $\pi_\theta({\bf s}_{-1}) = \bm a_0$ for a fixed initial DM commands $ \bm a_0$. In particular, we focus here on parametric models of $\pi_\theta$ where $\theta$ is the set of parameters of the policy, e.g., the weights and biases of a neural network. That is, given that the actions are given by $\pi_\theta$, we wish to find the parameters $\theta$ that maximize the expected cumulative reward the decision-maker receives. Here $T$ is the maximum length of an \emph{episode} or a single run of the algorithm in the environment.

The transition dynamics is usually not known in adaptive optics control: it includes a multitude of unknowns including the atmosphere turbulence, dynamics of the WFS and DM, and the jitter in the computational delay. In order to solve Eq. \eqref{eq:obj_of_rl} efficiently, \emph{model-based} RL algorithms estimate the true dynamics model $p(s_{t+1}|s_t, a_t)$ in \eqref{eq:obj_of_rl} by an approximate model $\hat{p}(s_{t+1}|s_t, a_t)$. \emph{Model-free} methods, in turn, only learn a policy -- they do not attempt to model the environment.

The standard MDP formulation assumes that all information about the environment is contained in the state $s_t$. This is not the case in many real-world domains, such as adaptive optics control. A more refined formulation is then the \emph{partially observed} MDP or POMDP, where the decision-maker observes $o_t$, which is some subset or function of the true underlying state. Note that the Markov property, i.e., the assumption that the next state depends only on the previous state and action, does not apply to the observations in a POMDP. This work uses the standard method of having our state representation include a small number of past observations (WFS measurements) and actions (control voltages) to deal with this issue. This allows the policy to use knowledge of past actions to predict the next action. The exact form of the observations $o_t$ and the full state $s_t$ for adaptive optics control will be given in Section \ref{sec:state_repre}.

Finally, it is common in reinforcement learning to use reward functions that are not differentiable (such as 1 for winning a game, 0 otherwise) or functions that do not depend directly on the state. In high-contrast imaging, we would like to minimize the speckle intensity in the post-coronagraphic PSF. However, this can be difficult to estimate at the high frequencies of modern HCI instruments. We discuss the specific choices in this regard in Section \ref{sec:state_repre}.



\section{Adaptive optics control}
\label{sec:ao}
This section introduces AO control aspects that are relevant to our work. First, we introduce the AO system components and then outline a standard control law called the \emph{integrator} and the related calibration process. An overview of the AO control loop is given in Figure \ref{fig:overview}; the incoming light $\phi_t^{tur}$ at the timestep $t$ gets corrected by the DM. Next, the WFS measures the DM corrected residual wavefront $\phi_t^{res}$. After receiving the wavefront sensor measurement, the control computer calculates a set of control voltages and sends the commands to the DM.

Further, the AO control loop inherits a temporal delay. The delay consists of a measurement delay introduced by the WFS integration and a control delay consisting of WFS readout, computation of the correction signal by the control algorithm, and its application to the DM. These add up to a total delay of at least twice the operating frame-time of the AO system \citep{madec1999control}. 

\subsection{PWFS for AO}\label{sec:pwfs}

The function of the WFS is to measure the spatial shape of the residual phase of a wavefront $\phi_t^{res}$. There are several different types of WFSs, but in this work, we focus on the so-called pyramid WFS (PWFS), which is a mature concept providing excellent performance for HCI \citep{guyon2005limits}. In the following, we give a short description of the PWFS.

The PWFS can be viewed as a generalization of the Foucault knife-edge test \citep{ragazzoni1996pupil}. In pyramid wavefront sensing, the electric field of the incoming wavefront is directed to a transparent four-sided pyramid prism. The prism is located in the focal plane of an optical system and, hence, can be modeled as a spatial Fourier filter that introduces specific phase changes according to the shape of the prism \citep{fauvarque2017general}. This four-sided pyramid divides the incoming light into four different directions, and most of the light is propagated to four intensity images on the PWFS detector. Due to the slightly different optical paths of the light, the intensity fields differ from each other. These differences are then used as the data for recovering the disturbances in the incoming phase screen.

Commonly, pyramid data, i.e., the intensity fields, are processed to so-called slopes $w_x, w_y$ that correlate positively to actual gradients fields of the phase screen. In this paper, we follow the approach of \cite{verinaud2004nature}, where the slopes are normalized with the global intensity. In practice, we receive a vector $\boldsymbol{w}$ that is a collection of the measurements $w_x, w_y$ at all possible locations $x, y$.

Both modulated and non-modulated Pyramid sensor observations are connected to the incoming wavefront via a non-linear mathematical model. This study considers non-modulated PWFSs, where the non-linearity is stronger, but the sensitivity is better at all spatial frequencies \citep{guyon2005limits}. Currently, most wavefront reconstruction algorithms utilize a linearisation of this model, inducing a trade-off between sensitivity and robustness (modulated PWFS vs. non-modulated PWFS). Machine learning techniques have the potential to overcome this trade-off and increase PWFS performance without a decisive robustness penalty.

Another feature of the PWFS is that its sensitivity varies depending on both the seeing conditions and the level of AO correction itself \citep{korkiakoski2008improving} which is mainly introduced by high spatial frequency aberrations which the DM cannot control. The presence of these aberrations reduces the signal strength of the measurement also for the controlled modes, and the strength of the reduction depends on the mode's spatial frequencies \citep{korkiakoski2008improving}.

\begin{figure}
\centering
\includegraphics[ clip,width=0.45\textwidth]{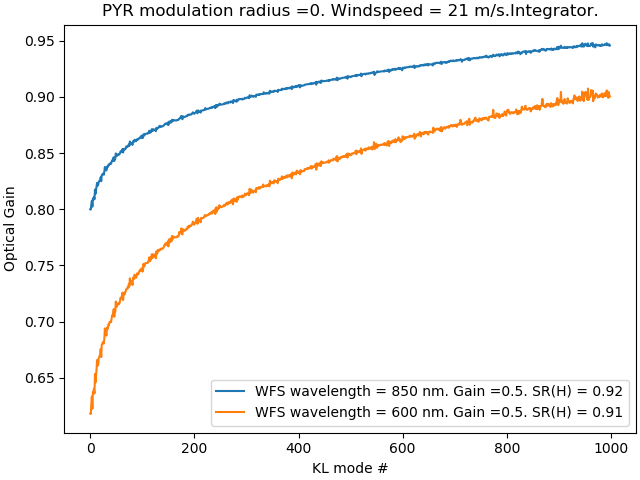}%
\caption{Modal optical gains for the case of an 8-m telescope with zero-modulation and integrator control and considering two different wavefront sensor wavelengths.} 
\label{fig:og1}
\end{figure}

To illustrate the OG effect of the Pyramid sensor, we use a preliminary version of a semi-analytical model code-named 'AO cockpyt' (in preparation). This model is based on the work of \cite{2019JOSAA..36.1241F}, describing the sensitivity of the Pyramid sensor in the presence of residuals, and on an adaptation of Fourier models from \cite{2010JEOS....5E0055J} and \cite{correia2020performance}. Figure \ref{fig:og1} shows the analytically predicted modal optical gains for the case of an 8-m telescope with zero-modulation and integrator control and considering two different wavefront sensor wavelengths. The assumed AO system for this analytical prediction is the same as the one used for our numerical simulations presented in Section 6 (41 x 41 actuators correct for seeing of 0.7" at 550nm at 1000 Hz framerate using a 0th magnitude guide star). The figure shows how the optical gain depends on the spatial frequency of the control modes (the K-L are numbered from low- to high spatial frequencies) and on the WFS Strehl ratio, which is lower at the shorter wavelength.

A modal optimization of the controller gains using the knowledge of Figure \ref{fig:og1} can solve most of the problems (diagonality assumption in \citet{2020A&A...644A...6C}) and applying the usual control theory margins (gain and phase) for ensuring a robust system. Determining optical gains in real-time is possible but complex (\cite{2021A&A...650A..41D}, \cite{2020A&A...644A...6C}), and the relative variations shown in Figure 2 are of the order 10-20\% for our XAO case. Hence, compensation for the mode-dependent optical gains with a single integrator gain may lead to acceptable results. However, an aggressive static integrator gain could impair loop robustness when the correction improves, and the optical gains increase. Section 6 presents evidence that PO4AO takes the PWFS OG effect into account for improved performance. Further, modal gain compensation of OG is a solution that is expected to work in favorable cases, but still, the non-linearities after OG compensation will remain and can only be treated with non-linear methods as the one studied in this paper.

\subsection{Classical AO control}
\label{sec:baseline}
Classically, an AO system is controlled by combining a linear reconstructor with a proportional-integral (PI) control law. We call this controller \emph{the integrator} and use it as the reference method for the comparison with PO4AO. As a starting point, the controller assumes to operate in a regime where the dependence between WFS measurements and DM commands is linear to a good approximation, satisfying
 \begin{equation}
    \label{eq:ip}
    \bm w_t = D v_t + \xi_t,   
\end{equation}
where $\bm w_t = (\delta w^1_t, \cdots, \delta w^n_t)$ is the WFS data, $v_t$ the DM commands and $D$ is so-called \emph{interaction matrix}. Moreover, $\xi_t$ models the measurement noise typically composed of photon and detector noise. The DM command vector $v_t$ represents the DM shape given in the function subspace linearly spanned by the DM influence functions. 

The interaction matrix $D$ represents how the WFS sees each DM command. It can be derived mathematically if we accurately know the system components (WFS and DM) and the alignment of the system. In practice, it is usually measured by poking the DM actuators with a small amplitude staying inside the linear range of the WFS, and recording the corresponding WFS measurements \citep{2004JOSAA..21.1004K, 2021MNRAS.501.3443L}.


The interaction matrix $D$ is generally ill-conditioned, and regularization methods must be used to invert it \citep{engl1996regularization}. Here, we regularize the problem by projecting $v_t$ to a smaller dimensional subspace spanned by Karhunen--Lo\'{e}ve (KL) modal basis. The KL basis is computed via a double diagonalization process, which considers the geometrical and statistical properties of the telescopes \citep{1994ESOC...48..187G}. This process results in a transformation matrix $Bm$ which maps DM actuator voltages to modal coefficients.

We observe that the \emph{modal interaction matrix} is now obtained as $DB_m^\dagger$, where $B_m^\dagger$ is the Moore--Penrose pseudo-inverse of $B_m$.  A well-posed \emph{reconstruction matrix} for the inverse problem in \eqref{eq:ip} is then given by
\begin{equation}
C_m = (D P_m)^\dagger,
\end{equation}
where $P_m = B_m^\dagger B_m$ is a projection map to the KL basis. Regularization by projection is a classical regularization with well-established theory \cite{engl1996regularization}. It is well-suited for the problem at hand due to the physics-motivated basis expansion and fixed finite dimension of the observational data.

With $\Delta \bm w_t$ denoting the residual error seen by the WFS in closed loop, and t denoting the discrete time step of the controller, the integrator control law is

\begin{equation}
\Tilde{\bm v}^t = \Tilde{\bm v}^{t-1} + g C \Delta \bm w^t,
\end{equation}
where $g$ is so-called \emph{integrator gain}. In literature, $g<0.5$ is typically found to provide stable control for a two-step delay system \cite{madec1999control}.

\section{Learning to control using a model}
Here we detail the control algorithm including optimization for the dynamics model $p_\omega(\st, \action)$ and the policy $\pi_\theta(\action | \st)$. In standard AO terms, the policy combines the reconstruction and control law (e.g., a least-squares modal reconstruction followed by integrator control); in our case, a non-linear correction to a least-squares modal reconstruction (MDP formulation) and a predictive control law. The key idea is to learn a dynamics model that predicts the next wavefront sensor measurement given the previous measurements and actions and to use that model to optimize the policy. Our method iterates the following three phases: \footnote{See Algorithm \ref{alg:mbpo} for more details}
\begin{enumerate}
    \item \textbf{Running the policy:} we run the policy in the AO control loop for $T$ timesteps (a single episode).
    \item  \textbf{Improving the dynamics model:} we optimize the dynamics model using a supervised learning objective Eq. \eqref{eq:dynamics_error}.
    \item  \textbf{Improving the policy:} We optimize the policy using the dynamics model \eqref{eq:optimization_of_hatr}.
\end{enumerate}

At each iteration of our algorithm, we collect an episode's worth of data, e.g., 500 subsequent sensor measurements and mirror commands, by running the policy in the AO control loop for $T$ timesteps. We then save the observed data and given actions and train our policy and dynamics model using gradients computed from all previously observed data. 

The following sections discuss how we represent each observation, our convolutional neural network architecture for both the dynamics model and the policy, and the optimization algorithm itself.  

\subsection{Adaptive optics as an MDP}
\label{sec:state_repre}
We define the adaptive optics control problem as an MDP by following the approach of \cite{nousiainen2021adaptive}. As discussed in Sections \ref{sec:prelim} and \ref{sec:ao}, we do not directly observe the state of the system but instead observe a noisy WFS measurement. In addition, adaptive optics systems suffer from \emph{control delay} resulting from the high speed of operation, which means that the system evolves before the latest action has been fully executed. Hence, we set our state presentation to include a small amount of past WFS measurements and control voltages.

We denote the control voltages applied to DM at a given time instance $t$ by  $\Tilde{\bm v}_{t}$ and the pre-processed PWFS measurements by $\bm{w}_{t}$. We define the set of actions to be the set of differential control voltages:

\begin{equation}
\bm a_t = \Delta \Tilde{\bm v}_{t}.
\end{equation}

In adaptive optics, at each timestep $t$, we observe the wavefront sensor measurement $\bm w_t$. We project the measurement into voltage space by utilizing the reconstruction matrix $C$. The observation is then given by the quantity:
\begin{equation}
    \obs = C \bm w_t.
\end{equation}
To represent each state, we concatenate previous observations and actions. That is,
\begin{equation}
\bm s_t =  \begin{pmatrix} \bm o_{t}, \bm o_{t-1}, \dots, \bm o_{t-k}, \bm a_{t-1}, \bm a_{t-2}, \dots, \bm a_{t-m} \end{pmatrix},
\end{equation}
where we choose $k = m$ (as in the typical pseudo-open-loop prediction). The state includes data from the previous $m$ time steps and the reconstruction matrix $C$. Here the reconstruction matrix serves solely as a pre-processing step for WFS measurements. It speeds up the learning process by simplifying the convolutional NN (CNN) architecture (same dimensional observations and actions). However, It does not directly connect the measurement to actions and, therefore, using it does not imply a sensitivity to misregistration \citep{nousiainen2021adaptive}.  

For a state-action pair, the reward is chosen as the residual voltages' negative squared norm corresponding to the following measurement:
\begin{equation}
r(\bm s_t, \bm a_t) = -{\mathbb E}_{p(\bm s_{t+1} | \bm s_t, \bm a_t)} \| \tilde{\bm o}_{t+1} \|^2,
\end{equation}
where $\tilde{\bm o}_{t+1}$ is obtained from $\tilde{\bm s}_{t+1} \sim p(\cdot| s_t, a_t)$.

This quantity is proportional to the observable part of the negative norm of the true residual wavefront. This reward function does not capture all error terms such as aliasing and non-common path errors (NCPA), and hence, the final contrast performance will always be limited by these. The aliasing errors could be mitigated with traditional means, e.g., by introducing a spatial filter \citep{poyneer2004spatially} or by oversampling the wavefront, i.e., by using a WFS with finer sampling than the one provided by the DM. We also already eluded on the fact that minimizing the residual wavefront seen by the WFS does not necessarily minimize the residual halon in the science image because of NCPA between the two. PO4AO could treat NCPA by including science camera images in the state formulation, but these would have to be provided at the same cadence as the WFS data, which is usually not the case. Still, NCPA can be handled by PO4AO in the usual way by offsetting the WFS measurements by an amount determined by an auxiliary image processing algorithm (e.g., \cite{give2007broadband}, \cite{paul2013coronagraphic}. Finally, the reward does not include an assumption on the time delay of the system, so the method learns to compensate for any delay and predict the wavefront.

\subsection{The dynamics model}
\label{sec:spatial}

\begin{figure*}[htp]
    \centering
    \includegraphics[trim={0cm 11cm 10cm 1cm}, clip,width=0.95\textwidth]{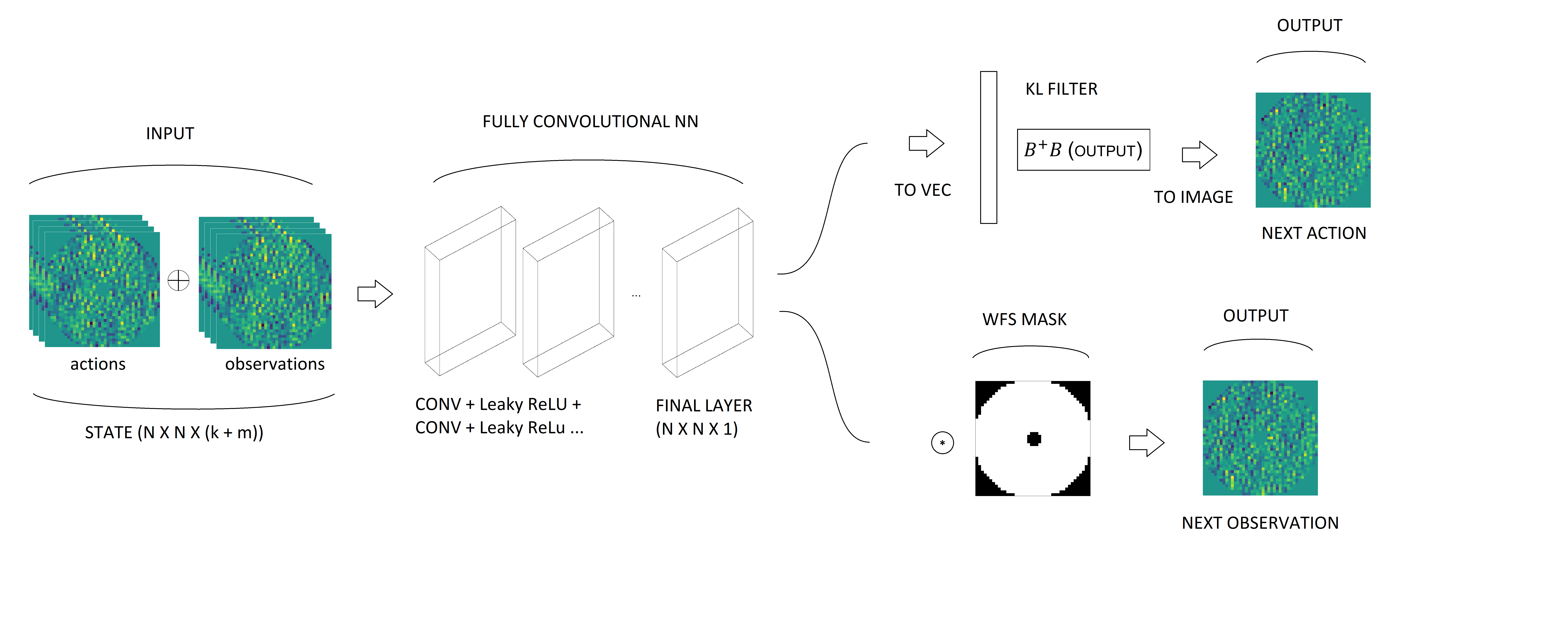}
    \caption{Neural network architectures. Both the dynamics model and the policy NN take same input: concatenations of past actions and observations. They also share the same fully convolutional structure in the first layers. At the output layer, the policy model includes the KL-filtering scheme (upper right corner) and the dynamics model output is multiplied with the WFS mask (lower right corner). See Section \ref{sec:spatial} for details.}
    \label{fig:models}
\end{figure*}

An adaptive optics system inherits strong spatial correlations in observations and control space - neighboring actuators and WFS pixels close to each are more correlated than actuators further apart due to the steep negative slope of the turbulence temporal PSD \citep{fried1990time} and the frozen flow hypothesis. We employ a standard fully convolutional neural network (CNN), equipped with a leaky rectified linear unit (LReLU, \cite{Maas2013RectifierNI}) activation functions that predicts the next wavefront sensor readout. The CNN should work well for our setup with DM actuators and WFS subapertures aligned on a grid in a spatially homogeneous geometry.

In practice, the state is a 3D tensor (matrices stack along the third dimension, i.e., a $(N \times N \times (k+m))$ tensor) with the channel dimension corresponding to DM actuator grid (2D) and the number of previous observations (k) and actions (m). See Figure \ref{fig:models} for an illustration. 

The deterministic dynamics model $\hat{p}_{\omega}(\st, \action)$ estimates the next state $\stp$ given the previous state and action. The model parameters $\omega$ (i.e., the NN weights and biases) are trained by first running the policy $\pi$ in the environment, i.e., controlling the AO system with the policy, collecting tuples of $(\st, \action, \stp)$ into a dataset $\mathcal{D}$, and minimizing the squared difference between the true next states and the predictions 

\begin{equation}
\label{eq:dynamics_error}
    \sum_{\mathcal{D}} \left\|\stp - \hat{p}_\omega(\st, \action)\right\|^2
    = \sum_{\mathcal{D}} \left\|\bm o_{t+1} - \hat{\bm o}_{t+1}\right\|^2,
\end{equation}
where ${\bm o}_{t+1}$ is obtained from the state $\stp$ and $\hat{\bm o}_{t+1}$ is the observation predicted by $\hat{p}_\omega(\st, \action)$. The optimization is done using the Adam algorithm \citep{kingma2014adam}. Again we do not assume any integer time delay here, but as the past actions are included in the state formulation, we learn to compensate for it.

It is well-known that model-based RL performance unfavorably exploits an overfitted dynamics model in the control (e.g., planning or policy optimization), especially in the early stages of training \citep{nagabandi2018neural}. To discourage this behavior, we employ an ensemble of several models, each of which is trained using different bootstrap datasets, i.e., subsets of the observations collected during training. In practice, this means that each model sees a different subset of observations, leading to different NN approximations. During policy training, predictions are averaged over the models (line 9 of Algorithm \ref{alg:mbpo}). See e.g \cite{chua2018deep} for a more detailed discussion on ensemble models.

\subsection{The policy model}
Again, we employ a fully convolutional neural network as the policy, similar to the dynamics model. The input is a 3D tensor representing the state, and the output is a 2D tensor (a matrix) representing the actuator voltages. The WFS measurement is blind or insensitive to some shapes of the mirror, such as the well-known waffle mode and actuators on the boundary. We ensure that we do not control these modes by projecting each set of control voltages to the control space, i.e., we reshape the 2D output to a vector, multiply it by a filter matrix, and then reshape the output back to a 2D image. The full policy model $\pi$ is given by

\begin{equation}
\pi_\theta(\st) =  B^{\dagger}B F_\theta(\st),    
\end{equation}
where $B^{\dagger}B$ projects the control voltages onto the control space defined by the K-L modes and $F_\theta$ is the standard fully convolutional NN, where the output is vectorized. Figure \ref{fig:models} gives more detailed overview of the network architecture of $F_\theta$.

\subsection{Policy optimization}\label{sec:PO4AO}

Ideally, the policy $\pi_\theta(\st)$ would be optimized based on the expected cumulative reward function \eqref{eq:obj_of_rl}. However, as we do not have access to the true dynamics model $p$, we must approximate it with the learned dynamics model $\hat p_\omega$.
To stabilize this process, we introduce an extended time horizon $H \ll T$ over which the performance is optimized. Let us define
\begin{equation}
    \hat r_\omega(\st,\action) = - \|\tilde {\bm o}_{t+1}\|^2,
\end{equation}
where $\tilde {\bm o}_{t+1}$ is obtained from $\tilde {\bm s}_{t+1} = \hat p_\omega(\st, \action)$.
This leads to the approximative policy optimization problem
\begin{equation}
    \label{eq:optimization_of_hatr}
    \argmax_\theta \sum_{{\bf s} \in \mathcal{D}}  \sum_{t=1}^{H} \hat r_\omega(\tilde \st, \pi_\theta(\tilde \st)),
\end{equation}
where $H$ the planning horizon and
\begin{equation*}
    \tilde {\bf s}_1 = {\bf s} \quad \text{and} \quad \tilde {\bf s}_{t+1} = \hat p_\omega(\tilde \st, \pi_\theta(\tilde \st)).
\end{equation*}
Here the planning horizon $H$ is chosen based on the properties of the AO system. More precisely, for AO control, the choice of the planning horizon H is driven by the system's control delay. In the case of a simple two-frame delay, no DM dynamic, and no noise, we would plan to minimize the observed wavefront sensor measurements two steps into the future, i.e., we would implicitly predict the best control action by the DM at the time of the corresponding WFS measurement. However, the effective planning horizon is longer in the presence of DM dynamics and temporal jitter since the control voltage decisions are not entirely independent. The choice of the planning horizon is a compromise between two effects: too short a planning horizon jeopardizes the loop stability, and too long a planning horizon makes the method prone to overfitting. We use $H=4$ frames in all our experiments (numerical and laboratory) as a reasonably well-working compromise.

The policy $\pi$ is optimized by sampling initial states from previously observed samples, computing actions for them, and using the dynamics model to simulate what would happen if we were to take those actions. We can then use the differentiable nature of both our models and the reward function to backpropagate through rewards computed at each timestep. More specifically, at each iteration, we sample a batch of initial states $s_\tau$ and compute the following $H$ states using the dynamics model. We then have $H$ rewards for each initial state, and we use the gradients of the sum of those rewards with respect to the policy parameters $\theta$ to improve the parameters. The full procedure of training the dynamics and the policy is given in Algorithm \ref{alg:mbpo}, where the while-loop (line 3) iterates over episodes and lines 6 - 16 execute an update of policy via policy optimization. 

\begin{algorithm}
\caption{Policy Optimization for Adaptive Optics (PO4AO)}
\label{alg:mbpo}
\begin{algorithmic}[1]
\State Initialize policy and dynamics model parameters $\theta$ and $\omega$ randomly
\State Initialize gradient iteration length $K$, batch size $B<|\mathcal{D}|$ and planning horizon $H$
\While{not converged}
    \State Generate samples $\{s_{t+1}, s_t, a_t \}$  by running policy $\pi_\theta(a_t|s_t)$ for $T$ timesteps (an episode) and append to $\mathcal{D}$
  \State Fit dynamics by minimizing Eq. \eqref{eq:dynamics_error} w.r.t $\omega$ using Adam
  \For{iteration $k=1$ to $K$}
      \State Sample a mini batch of $B < |\mathcal{D}|$ states $\{ s_\tau \}$ from $\mathcal{D}$
      \For{each ${\bf s}_\tau$ in the mini batch}
        \State Set $\tilde {\bf s}_1^\tau = {\bf s}_\tau$
        \For{$t=1$ to $H$}
            \State Predict $\action = \pi_\theta(\st)$
            \State Predict $\stp = \hat{p}_\omega(\st, \action)$
            \State Calculate $R_t = \hat r_\omega(\st, \action)$
        \EndFor
      \EndFor
      \State Update $\theta$ by taking a gradient step according to $\nabla_\theta \sum_{t=\tau}^{ \tau + H} R_t $ with Adam.
  \EndFor
\EndWhile
\end{algorithmic}
\end{algorithm}

\begin{table*}[ht]
    \centering
    \caption{ Simulations parameters}
    \label{table:simulator_parameters2}
\begin{tabular}{ c c c} 
 \hline\hline
 \multicolumn{3}{c}{Telescope "VLT"} \\
 \hline
         Parameter  & Value  &  Units  \\
 \hline
 Telescope diameter   &  8  & m     \\
 Obstruction ratio    &  14  & percent           \\
 Sampling frequency   &  1000  & Hz        \\
 Active actuators     &  1364   & actuators         \\
 PWFS subapertures     & $41\times41$ & apertures           \\
 PWFS modulation           & 0        &  $\lambda$ / D\   \\
 Photon flux 0/9 mag &   $1.25 \times 10^8$/$3.1 \times 10^4
$   &   photons / frame / aperture          \\ 
 DM coupling    &   0.3 & percent \\
 DM influence functions    &  "Gaussian"  & $\cdots$   \\
 WFS wavelength & 0.85 &  \textmu  m  \\
 Science camera wavelength & 1.65 & \textmu  m \\
 \hline
 \multicolumn{3}{c}{Telescope "ELT"} \\
 \hline

 Telescope diameter   &  40  & m     \\
 Obstruction ratio    &  30  & percent           \\
 Sampling frequency   &  1000  & Hz        \\
 Active actuators     &  10556   & actuators         \\
 PWFS subapertures     & $121\times 121$ & apertures           \\
 PWFS modulation           &  2          &  $\lambda$ / D\\  
 Photon flux 0th mag &   $2.7 \times 10^9$   &   photons / frame / aperture          \\ 
 DM coupling    &   0.3 & percent \\
 DM influence functions    &   "Gaussian" & $\cdots$ \\
 WFS wavelength & 0.85 &  \textmu  m  \\
 Science camera wavelength & 1.65 &  \textmu  m  \\
  \hline
  
 \multicolumn{3}{c}{Atmosphere parameters} \\
 \hline
 Fried parameter     &  16  & cm @ 500 nm     \\
 Number of layers    &  3  & $\cdots$       \\
 Layer altitudes    &  0 / 4 / 10  & km       \\
 $C_N^2$  & 50 / 35 / 15  & percent (\%)        \\
 Wind speeds   &  10 / 26 / 35  & m/s      \\
 Wind directions       & 0 / 45 / 180   &   degrees           \\    
 $L_0$ ($m$)      & 30 / 30 / 30   &   m        \\ 
  \hline
 \multicolumn{3}{c}{PO4AO parameters} \\
 \hline

 Planning horizon (H)     &  4  & steps     \\
 Past DM commands (m)     &  15  & commands       \\
 Past WFS measurements (k)  & 15  & frames        \\
 CNN ensemble size        & 5   &    $\cdots$          \\   
 Dynamics iterations / episode & 15 & steps\\
 Policy iterations / episode & 10 & steps \\
 Training mini batch size  & 32 &  $\cdots$ \\
  \hline

 \multicolumn{3}{c}{Fixed CNN parameters} \\
 \hline
 Number of conv. layers     &  3  & layers     \\
 Filter size     &  $3 \times 3$  & pixels       \\
 Padding  & 1  & pixels        \\
 Activation functions        &  Leaky ReLU   &       $\cdots$       \\             
  \hline
  
\end{tabular}

\end{table*}

\section{Numerical simulations}

\begin{table*}
\caption{Performance of 11 different 3-layer CNNs. All CNN models were trained from scratch with the same PO4AO parameters (see Table \ref{table:simulator_parameters2}) and VLT 0-mag simulation environment (see Section \ref{sec:setup-d} and Table \ref{table:simulator_parameters2}). The Strehl and reward were calculated from the last 1000 steps of the experiment. The inference time was also calculated for VLT and ELT-scale systems, while the training time after each episode was only calculated for the VLT-scale system due to computational limitations. The corresponding integrator performance (dominated by the fitting and temporal error) for the "VLT" simulation was 93.59 / -10 085 (Strehl/Reward).}              
\label{table:3}      
\centering                                      
\begin{tabular}{c c c c c c} 
 \hline\hline
 \multicolumn{6}{c}{CNN design} \\
 \hline
          & Filters & Past frames (k \& m) & Inf. speed (VLT/ELT) & Tr. time / episode (VLT) & Strehl/reward (VLT 0-mag) \\  
 \hline                                   
    CNN 1 & 32 & 10 & 0.29 / 0.35 ms & 1.4 sec & 95.61 / -4101 \\  
    \textbf{CNN 2} & 32 & 15 & 0.30 / 0.37 ms & 1.5 / 7 (ELT) sec & 95.69 / -3340 \\ 
    CNN 3 & 32 & 20 & 0.30 / 0.40 ms & 1.6 sec & 95.74 / -3029 \\ %
    CNN 4 & 32 & 25 & 0.30 / 0.43 ms & 1.8 sec & 95.75 / -2934 \\ 
\hline
    CNN 5 & 64 & 10 & 0.30 / 0.67 ms & 2.0 sec & 95.60 / -4002 \\ 
    CNN 8 & 64 & 15 & 0.31 / 0.70 ms & 2.2 sec & 95.75 / -3253 \\ 
    CNN 7 & 64 & 20 & 0.31 / 0.74 ms & 2.5 sec & 95.75 / -3052 \\ 
    CNN 8 & 64 & 25 & 0.32 / 0.79 ms & 2.5 sec & 95.76 / -2845  \\ 
\hline  
    CNN 9  & 128 & 10 & 0.36 / 1.52 ms & 3.7 sec & 95.65 / -3656\\ 
    CNN 10 & 128 & 15 & 0.37 / 1.58 ms & 3.8 sec & 95.71 / -2943 \\ 
    CNN 11 & 128 & 20 & 0.38 / 1.63 ms & 4.7 sec & 95.76 / -2847\\ 
\end{tabular}
\end{table*}

\subsection{Setup description}\label{sec:setup-d}
We evaluate the performance of PO4AO by numerical simulations. We use the COMPASS package \citep{ferreira2018compass} to simulate an XAO system at an 8-m employing a non-modulated Pyramid WFS in low noise (0 mag) and moderately large noise (9 mag) conditions. For comparison, we also consider the theoretical case of an "ideal" wavefront sensor where the wavefront reconstruction is simply a projection of the 2D-turbulence screen onto the DM's influence functions.

We also include a simulation of a 40-meter telescope XAO with PWFS to confirm that PO4AO nicely scales with aperture size and XAO degrees of freedom. Comprehensive error analysis and fine-tuning are left for future work. In order to stabilize the performance of the integrator, we added $2\lambda/D$ modulations to the PWFS. 

For all simulations, we simulate the Atmospheric turbulence as a sum of three frozen flow layers with Von Karman power spectra combining for Fried parameter $r0$ of $16$ cm at $500$ nm wavelength. The complete set of simulation parameters is provided in Table 1. 

We compare PO4AO against a well-tuned integrator and instantaneous controller, not affected by measurement noise or temporal error. For the Pyramid WFS, it still propagates aliasing and the fitting error introduced by uncontrolled or high-spatial frequency modes. For the idealized WFS, it acts as a spatial high-pass filter, instantaneously subtracting the turbulent phase projected on the DM control space (DM fitting error only).

In particular, the simulation setups were chosen to demonstrate the following key properties of the proposed method:

\begin{enumerate}
    \item The method achieves the required real-time control speed while being quick to train. This property enables the controller to be trained just before the science operation and be further updated during the operation. Consequently, the method is trained with the most relevant data and does not need to generalize to all possible conditions at once. Further, the method retains these properties with an ELT-scale instrument.
    \item The method is a predictive controller, robust to non-linear wavefront sensing and photon noise.
    \item The method can cope with the optical gain effect of the pyramid sensor. 
\end{enumerate}

\subsection{Algorithm setup}\label{algsetup}
We choose the state $\st$ (in MDP) to consist of $15$ latest observations and actions and set the CNN (dynamics and policy) to have 3-layers with 32 filters each. For further details on these choices, see Section \ref{sec:CNN}. The episode length is set to $500$ frames.

Each simulation starts with the calibrations of the system and the deriving of the reconstruction matrix $C$ and the K-L basis $B$; see Section 4. Note that the reconstruction matrix C serves solely as a filter that projects WFS measurement to control space. It does not have to match the actual registration of DM and WFS \citep{nousiainen2021adaptive}. In particular, the reconstruction matrix is measured around the null point in the calibrations and, hence, it suffers from the optical gain effect \cite{korkiakoski2008improving}. For PWFS simulations, the K-L filter set to include $85\%$ of total degrees of freedom, and for ideal wavefront sensing to filter matrix is an identity, i.e., no filtering included.

For all different conditions and instruments, we let simulations run until the performance of PO4AO is converged. That is $46000$ frames ($46$ seconds in real-time (theoretical)) with an episode length of $500$ frames. While the final contrast performance shown in Figures \ref{fig:0mag}, \ref{fig:9mag}, \ref{fig:data_mismatch} and \ref{fig:pcs_contrast} is calculated from the last $1000$ frames, we note that the correction performance very quickly passes the integrator performance as shown in Figure(s) \ref{fig:nowfs_train}, \ref{fig:training_0}, \ref{fig:training_9}, and \ref{fig:pcs_train}. After each episode, as described in Sections \ref{sec:PO4AO}, we halt the simulations and update the dynamics and policy models. Given the shallow convolutional structure (3 - layers and $32$ filters per layer) of the NN models and our moderate hardware, the combined (dynamic and policy) training time after each episode is about 1.5 seconds for VLT (and 7 seconds for ELT with the same training hyperparameters). For real-time implementation, training the NN models should be completed in the duration of an episode, i.e., in 0.5s ($500$ frames at 1 kHz). Given that we do not use the latest GPU hardware, and a NN update could also be done at a slower rate than after each episode, it is conceivable that this small gap can be overcome, and a real-time implementation of PO4AO is already possible.

The dynamics model can also be trained with data obtained with a different controller, e.g., the integrator or random control. Therefore, to improve the stability of the learning process, we \emph{warm up} the policy by running the first ten episodes with the integrator and added binary noise to develop a coarse understanding of the system dynamics:

\begin{equation}
\Tilde{\bm v}^t = \Tilde{\bm v}^{t-1} + g C \Delta \bm w^t + \sigma x,
\end{equation}
where x is binary noise, i.e., $-1$ or $1$ with the same probability, and $\sigma \in [0,1]$ is reduced linearly after each episode such that the first episode is run with high binary noise and the 10th episode with zero noise.

\subsection{CNN design and MDP state definition}\label{sec:CNN}
The PO4AO includes two learned models: the policy and the dynamics model. This paper aims to introduce an optimizations method called PO4AO to train the policy (from scratch) that minimizes the expected reward. The algorithm works for all differentiable function classes, e.g., neural networks. For simplicity, we choose to model the environment dynamics and policy using generic 3-layer fully convolutional neural networks. While further research is needed in finding the best possible architectures, we experimented with the number of convolutional filters per layer and the number of past telemetry data by testing the algorithm in the "VLT" environment with different combinations; see Table \ref{table:3}. We chose the model \textbf{CNN 2} to compromise between the overall performance, inference speed for VLT and ELT, and training speed. The chosen model performs well in all simulations and provides fast inference speed and fast training speed such that it could be completed during a single episode. Full model architecture optimization is left for future work (see Sections 8 for more details).

The inference speed in Table \ref{table:3} is the speed of the fully convolutional NN architecture inside the policy model (see Figure \ref{fig:models}). The total time control time includes two standard MVMs (pre-processing to voltages + KL filtering in the output layer) in addition to the inference time below. The inference time and training time were run with PyTorch on NVIDIA Quadro RTX 3000 GPU. Note here that given enough parallel computational power (e.g., GPU), the inference time of a fully convolutional NN is more determined by the number of layers and filter (same for VLT and ELT) than the input image's size. We observe that for CNN with fewer filters, the inference speed is very similar for VLT and ELT cases, while for heavier CNNs, the inference speed differs more with given hardware. The computational time of MVMs is naturally dependent on the DoF.

\subsection{Results}

\subsubsection{Training}
To evaluate the training speed of the method, we compare the learning curves (from which 5000 frames are obtained with the integrator + noise controller) of the method to the baseline of the integrator performance under the same realization of turbulence and noise (see Figs. \ref{fig:training_0}, \ref{fig:training_9}, \ref{fig:nowfs_train} and \ref{fig:pcs_train}). Since the simulations are computationally expensive, in the 40-meter telescope experiments, we compare the performance of the PO4AO only to average integrator performance (see Figure \ref{fig:pcs_train}).

We plotted the training curves with respect to total reward (the sum of normalized residual voltages computed from the WFS measurements) and Strehl ratio side by side. The method tries to maximize the reward, and consequently, it also maximizes the Strehl ratio. In all our simulations, the method achieves better performance than the integrator already after the integrator warm-up of $5000$ frames (5 sec on a real telescope), and the performance stabilizes at around $30000$ frames ($30$ sec). Since the fully convolutional NN structure can capture and utilize the homogeneous structure of the turbulence, the number of data frames needed for training of VLT and ELT control are on the same scale. However, training the same amount of gradient steps is computationally more expensive (although very parallelizable) for the ELT scale system.

\subsubsection{Prediction and noise robustness}
Here, we compare the fully converged PO4AO, the integrator, and ideal control in raw PSF contrast. We ran each controller for 1000 frames, and the wavefront residuals for each controller were propagated through a perfect coronagraph \citep{cavarroc2006fundamental}. The raw PSF contrast was calculated as the ratio between the peak intensity of non-coronagraphic PSF and the post-coronagraphic intensity field. A non-predictive control law suffers from the notorious wind-driven halo (WHD) \citep{cantalloube2018origin}, i.e., the butterfly-shaped contrast loss in the raw PSF contrast in Figures \ref{fig:nowfs}, \ref{fig:0mag}, \ref{fig:9mag}, \ref{fig:pcs_contrast}.

Figure \ref{fig:nowfs} assumes using the ideal WFS, i.e., the incoming phase is measured by a noiseless projection of the incoming phase onto the DM. Therefore, the ideal WFS eliminates aliasing and noise in the wavefront reconstruction process, only considering temporal and fitting errors. Further, we can easily eliminate temporal error in a simulation by directly subtracting the measured from the incoming phase.
The 'no noise, no temporal error' curve (black dashed) in Figure \ref{fig:nowfs} is therefore only limited by the ability of the DM to fit the incoming wavefront. The integrator with a 2-frame delay (blue curve) is then limited by the temporal error in addition. The PO4AO (red curve) largely reduces the WHD by predicting the temporal evolution of the wavefront but does not fully recover the fitting error limit (black dashed). Figure \ref{fig:nowfs}, therefore, demonstrates the ability of PO4AO to reduce the temporal error.

Figure \ref{fig:0mag} replaces the ideal WFS with the non-modulated PWS, which is affected by aliasing and requires some filtering of badly seen K-L modes during the reconstruction. Therefore, the 'no noise, no temporal error' contrast performance is worse than for the ideal WFS in Figure \ref{fig:nowfs}. The integrator with a 2-frame delay (blue curve) performs at a very similar contrast as in the ideal WFS case, so it is still limited mostly by temporal error. Again, PO4AO (red curve) lies about halfway between the integrator and 'no noise, no temporal error' controllers but performs at a reduced contrast compared to the ideal WFS case. Therefore, the PO4AO performance with the non-modulated PWS is affected by aliasing and reconstruction errors as well as the temporal error.

Figure \ref{fig:9mag} adds a significant amount of measurement noise. While this obviously does not affect the 'no noise, no temporal error' case, the contrast performance of both integrator and PO4AO is strongly reduced and dominated by noise. Still, PO4AO outperforms the integrator, which demonstrates the resilience of PO4AO against noise-dominated conditions. Finally, Figure \ref{fig:pcs_contrast} demonstrates that PO4AO maintains its properties in an ELT scale simulation.

Unfortunately, a "black box" controller like PO4AO does not allow us to cleanly separate all individual terms in the error budget because the controller's behavior is to some extent driven by the error terms themselves. However, as discussed above, we explored the relative importance of the individual terms by switching them on and off in our numerical experiments.

\subsubsection{Robustness against data mismatch}
So far, we have focused on static atmospheric conditions and size of the data set $\mathcal{D}$ is not limited, i.e., "ever-growing." However, in reality, the atmospheric conditions are constantly changing, creating a so-called data mismatch problem -- the prevailing atmospheric conditions are slightly different from the conditions in which the model was trained. To ensure the method's robustness to data mismatch, we train the model with very different conditions and then test the model with the original wind profile by plotting the raw PSF contrast averaged over $1000$ frames. We alter the wind by reducing the wind speed by 50 percent and adding $90$-degree variations to directions for training, i.e., we alter the spatial and temporal statistics of the atmosphere. We do not show the corresponding training plot since it was very similar to Figure \ref{fig:training_0}. The result of this experiment is shown in Figure \ref{fig:data_mismatch}. The integrator has naturally the same performance as before. The PO4AO still delivers better contrast close to the guide star but suffers from pronounced WDH further from the guide star. Most importantly, the PO4AO is robust and maintains acceptable performance even with heavy data mismatch, which could occur in the unlikely case that atmospheric conditions drastically change from one episode to the next, i.e., on a timescale of seconds. Anyhow PO4AO with limited data set size (old data irrelevant data removed) would adapt to such a change and recover the performance within the typical training times discussed in the previous paragraph.

\begin{figure}
\centering
\includegraphics[ clip,width=0.45\textwidth]{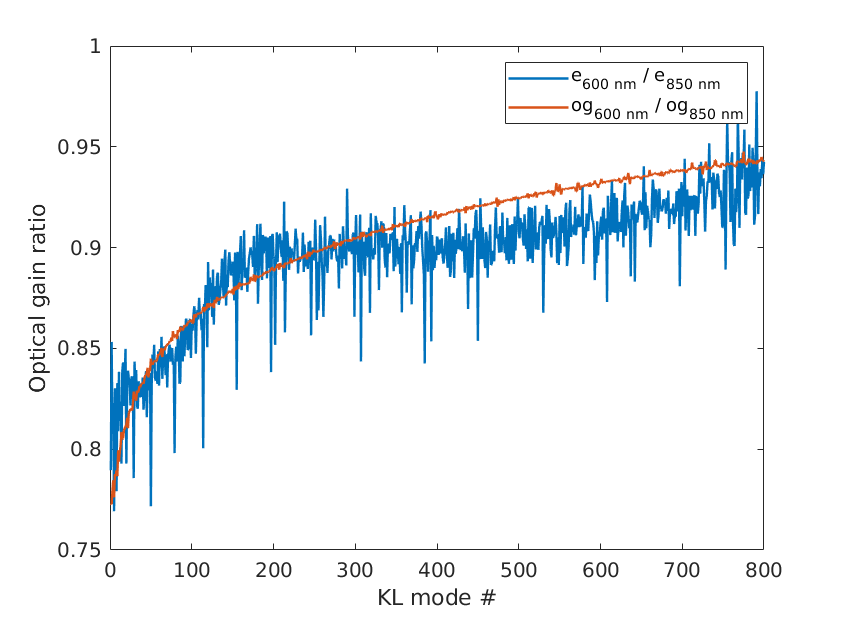}%
\caption{Sensitivity to the PWFS optical gain effect. The blue line corresponds to ratio between the optical gain estimates between the different wavelengths. The red line is the ratio between the semi-analytically derived optical gains at the two wavelengths (see Sections \ref{sec:pwfs} and Figure \ref{fig:og1}).} 
\label{fig:og}
\end{figure}

\subsection{Sensitivity to the PWFS optical gain effect}
PO4AO uses convolutional NNs and is, therefore, a non-linear method. Prospects are that it can adapt to non-linearities in the system, such as the optical gain effect observed for the Pyramid WFS. To examine this property, we run the following experiment. We control the non-modulated PWFS with PO4AO at 850 nm and 600 nm and record the policy after training. Then we control the PWFS with the integrator and record, in parallel, the actions PO4AO would have taken. The integrator control results in a correction performance similar to the Strehl ratios derived by the semi-analytical model (Figure \ref{fig:og1}). At the shorter wavelength, the PWFS sees larger residuals wavefront errors (in radian) and a stronger effect on the optical gains. However, if the controller can cope with such an effect, which we would expect for PO4AO, the suggested actions should counteract the dampened measurement. In order to validate this, we compare the ratios between the standard deviation of the observations (PWFS measurements) and the standard deviation of suggested PO4AO actions. We define an estimate for the optical gain compensation:

\begin{equation}
e_{\lambda} \propto \text{std}(\bm o_{int}^{\lambda})/ \text{std}(\bm a_{po4ao}^{\lambda}),
\end{equation}
where std is the temporal standard deviation, $\bm o_{int}^{\lambda}$ the observations while running the integrator, $\lambda$ the observing wavelength, and $\bm a_{po4ao}^{\lambda}$ the PO4AO suggested actions. As PO4AO is a predictive control method, this quantity also includes the effect of the prediction, i.e., it includes compensation for the temporal error as well. However, we can approximately cancel out the temporal error by comparing the ratio between optical gain estimates obtained at different wavelengths. The result of this experiment is shown in Figure \ref{fig:og}. We see that the empirical estimate for the optical gain sensitivity of PO4AO follows roughly the corresponding ratio of the two semi-analytically derived curves plotted in Figure \ref{fig:og1}. In particular, we see that the lower order modes are compensated more than high order modes. We, therefore, conclude that PO4AO adequately compensates for the optical gain effect of the PWFS.

\begin{figure*}
\centering
\subfloat[\label{fig:nowfs_train}]{%
  \includegraphics[trim={2.53cm 0.05cm 2cm 0.07cm}, clip,width=0.92\textwidth]{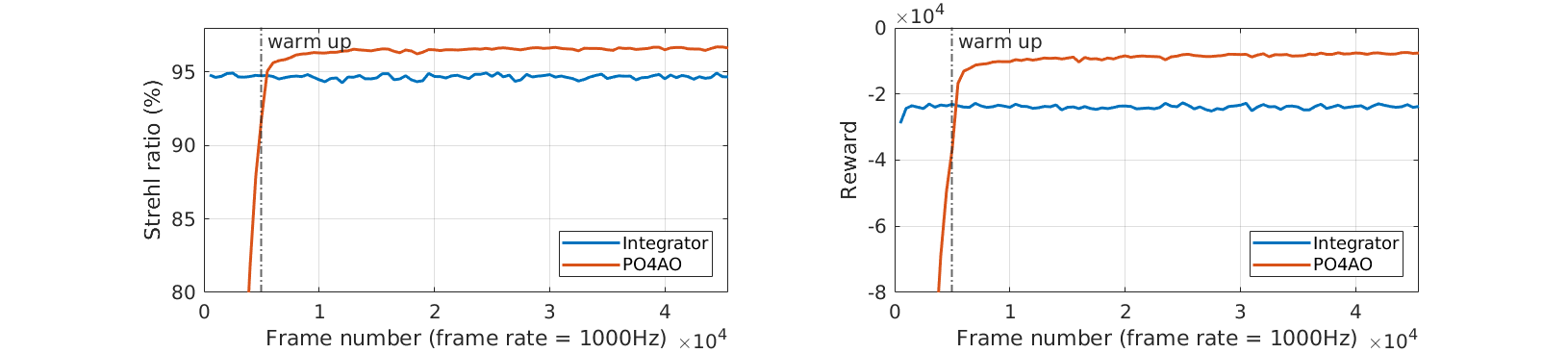}%
  }\par
\subfloat[\label{fig:training_0}]{%
  \includegraphics[trim={2.53cm 0.02cm 2cm 0.05cm}, clip,width=0.92\textwidth]{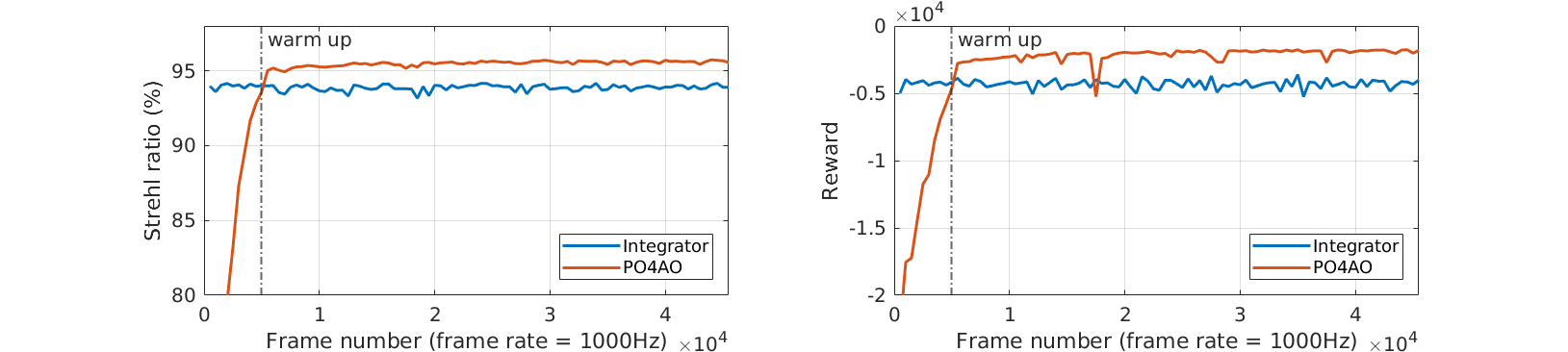}%
  }\par
\subfloat[\label{fig:training_9}]{%
   \includegraphics[trim={2.53cm 0.04cm 2cm 0.06cm}, clip,width=0.92\textwidth]{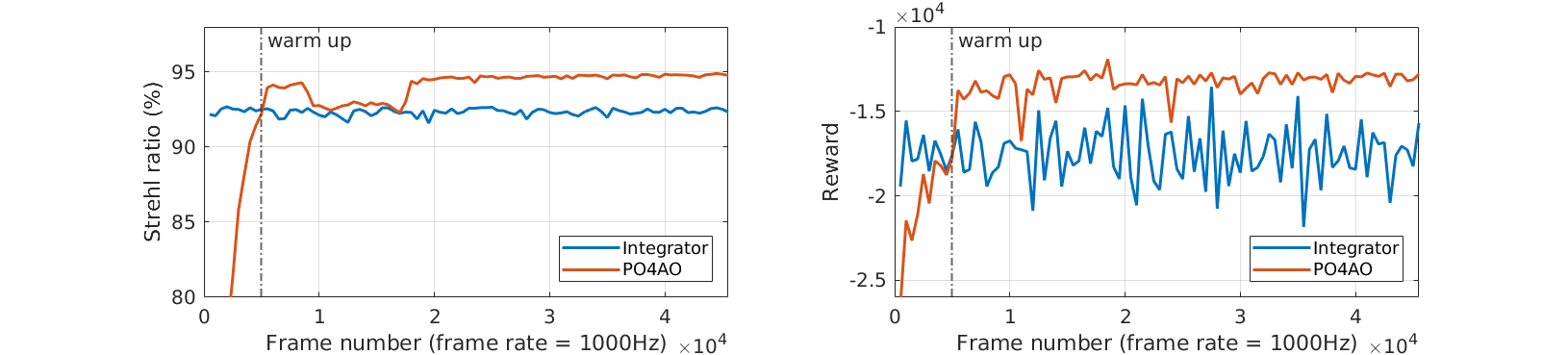}%
  }
  \caption{Training plots for 8-meter telescope experiments. Fig. \ref{fig:nowfs_train} is for ideal wavefront sensor, Fig. \ref{fig:training_0} is for the 0th magnitude guide star, and Fig. \ref{fig:training_9} for the 9th magnitude guide star. The red lines correspond to performance of the PO4AO during each episode and blue lines for the integrator. The gray dashed line marks the end of integrator warm up for PO4AO. In all cases the PO4AO outperforms the integrator all ready after the warm up period, in both the Strehl ratio and rewards. An optimized implementation of the PO4AO could run the training in parallel to control, and the training time would then be included in the plot (see Sec. \ref{algsetup}).}
\label{fig:training}
\end{figure*}

\begin{figure*}
\centering
\includegraphics[trim={2.8cm 0cm 2.5cm 0cm}, clip,width=0.92\textwidth]{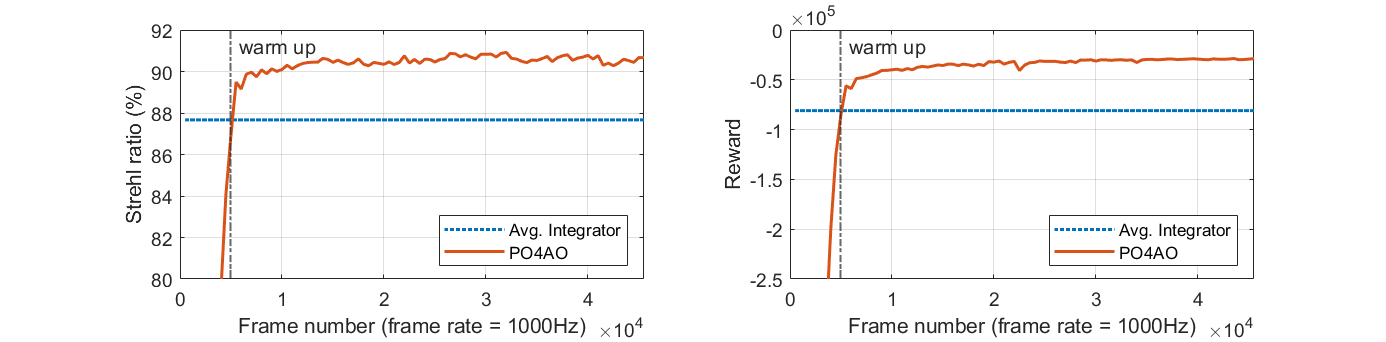}%
\caption{Training plots for the 40-meter telescope experiment. The red lines correspond to performance of the PO4AO during each episode and blue lines for the average integrator performance. The gray dashed line marks the end of integrator warm up for PO4AO. Similarly to 8-meter telescope experiments the PO4AO outperforms the integrator after the warm up.} 
\label{fig:pcs_train}
\end{figure*}

\begin{figure*}\centering
\subfloat[Performance of PO4AO with ideal WFS. We see that P04AO delivers a factor of 20-90 improvement inside the AO control radius compared to well-tuned integrator. \label{fig:nowfs}]
        {\includegraphics[trim={1.cm 0cm 2cm 0cm}, clip, width=0.48\textwidth]{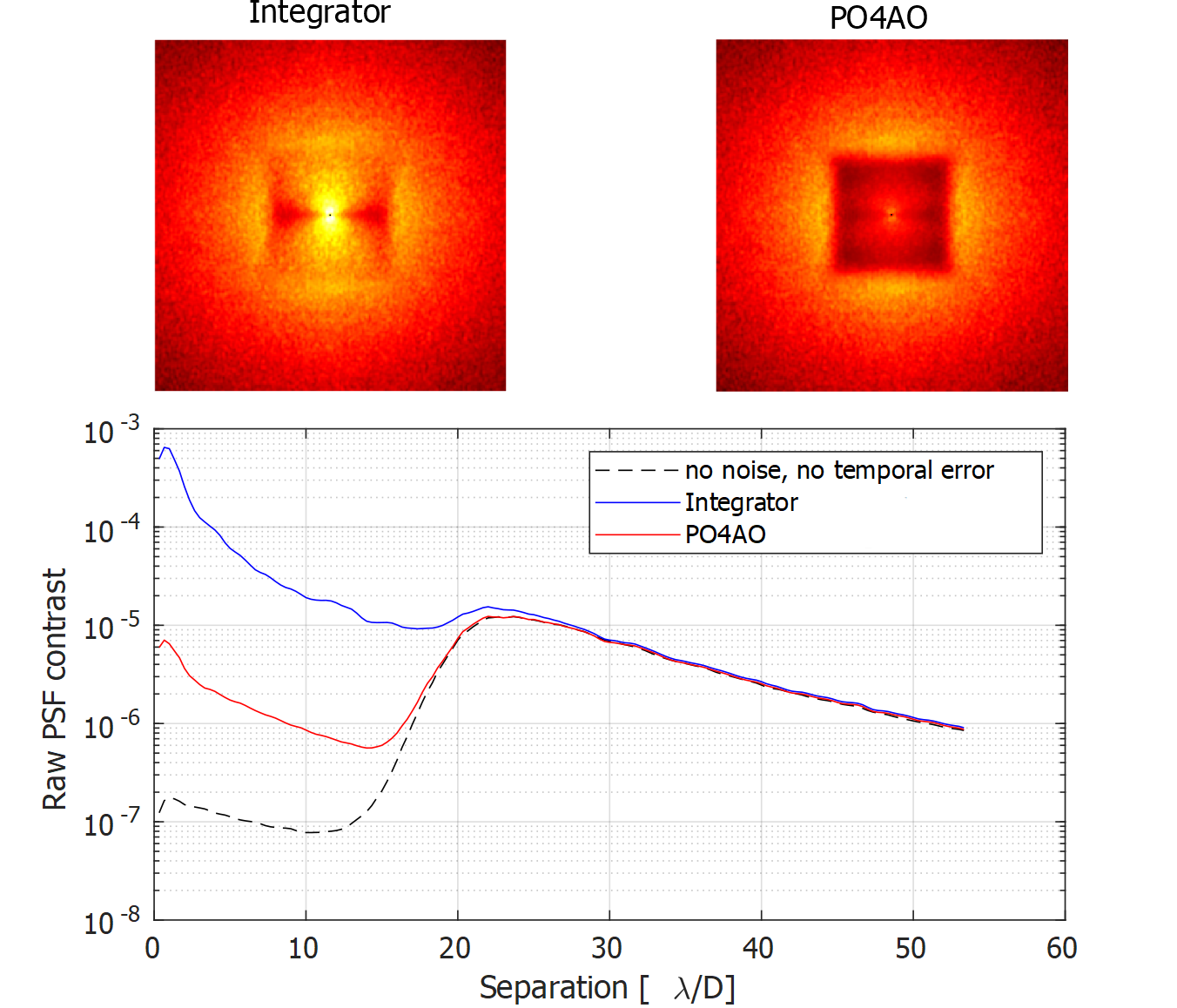}}
    \hfill
\subfloat[Performance of PO4AO on 0th mag guide star and a non-modulated PWFS. PO4AO delivers a factor of 4-7 better contrast inside the AO control radius.\label{fig:0mag}]
         {\includegraphics[trim={1.cm 0cm 2cm 0cm}, clip, width=0.48\textwidth]{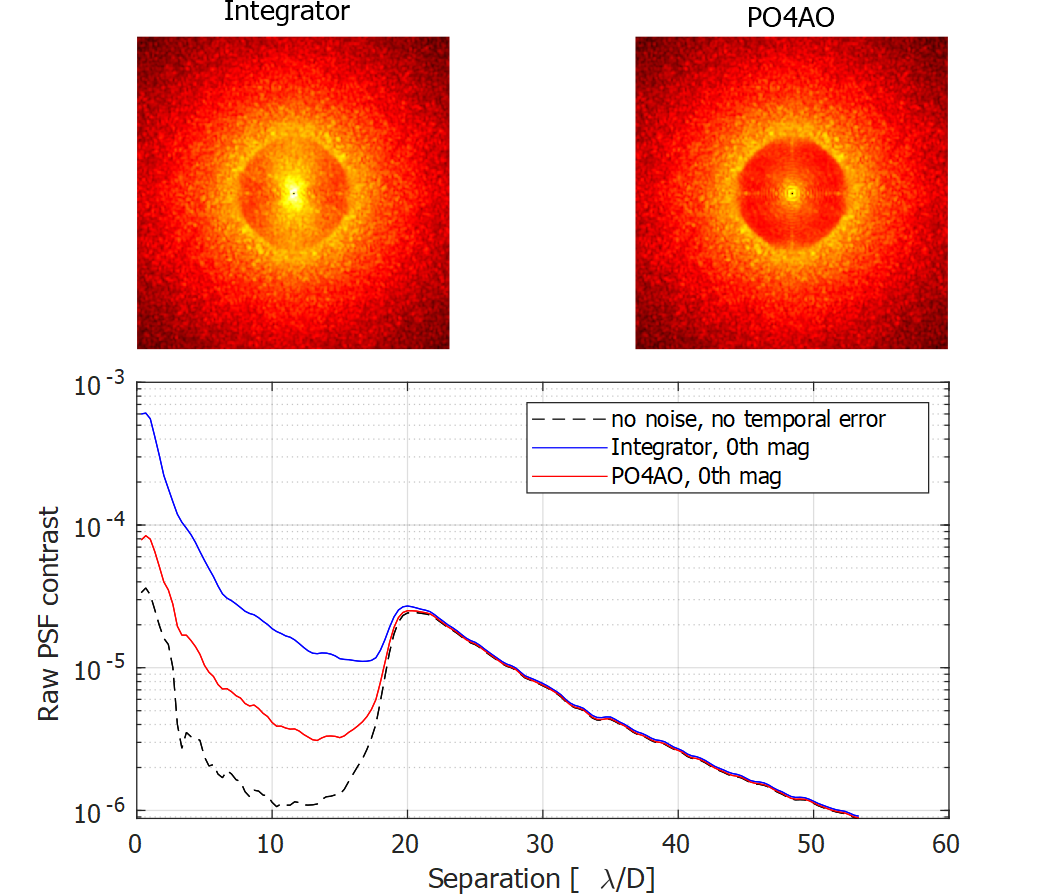}}

\subfloat[Performance of PO4AO on 9th mag guide star. We see a factor of 3-9 improvement in the raw PSF contrast. \label{fig:9mag}]
         {\includegraphics[trim={1.cm 0cm 2cm 0cm}, clip, width=0.48\textwidth]{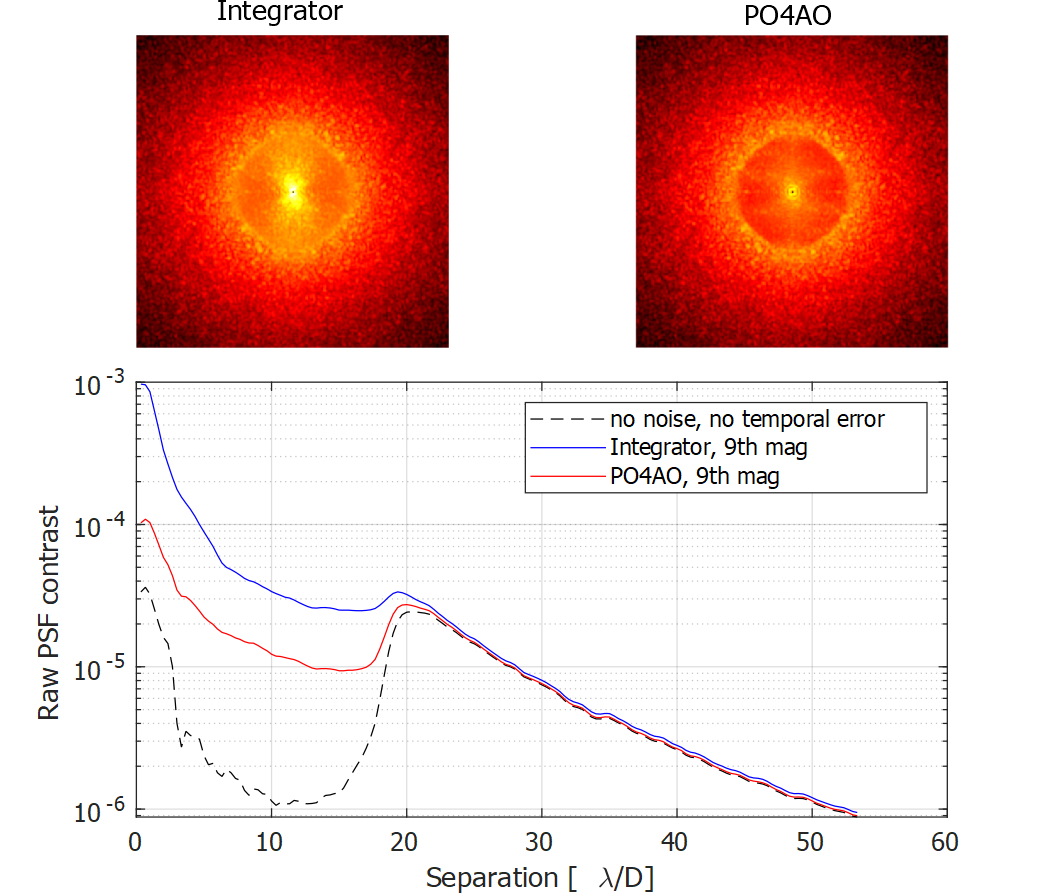}}
    \hfill
\subfloat[Performance of PO4AO under heavy data mismatch. PO4AO was trained with drastically different wind conditions. The PO4AO still delivers better contrast with small angular separations. \label{fig:data_mismatch}]
        {\includegraphics[trim={1.cm 0cm 2cm 0cm}, clip, width=0.495\textwidth]{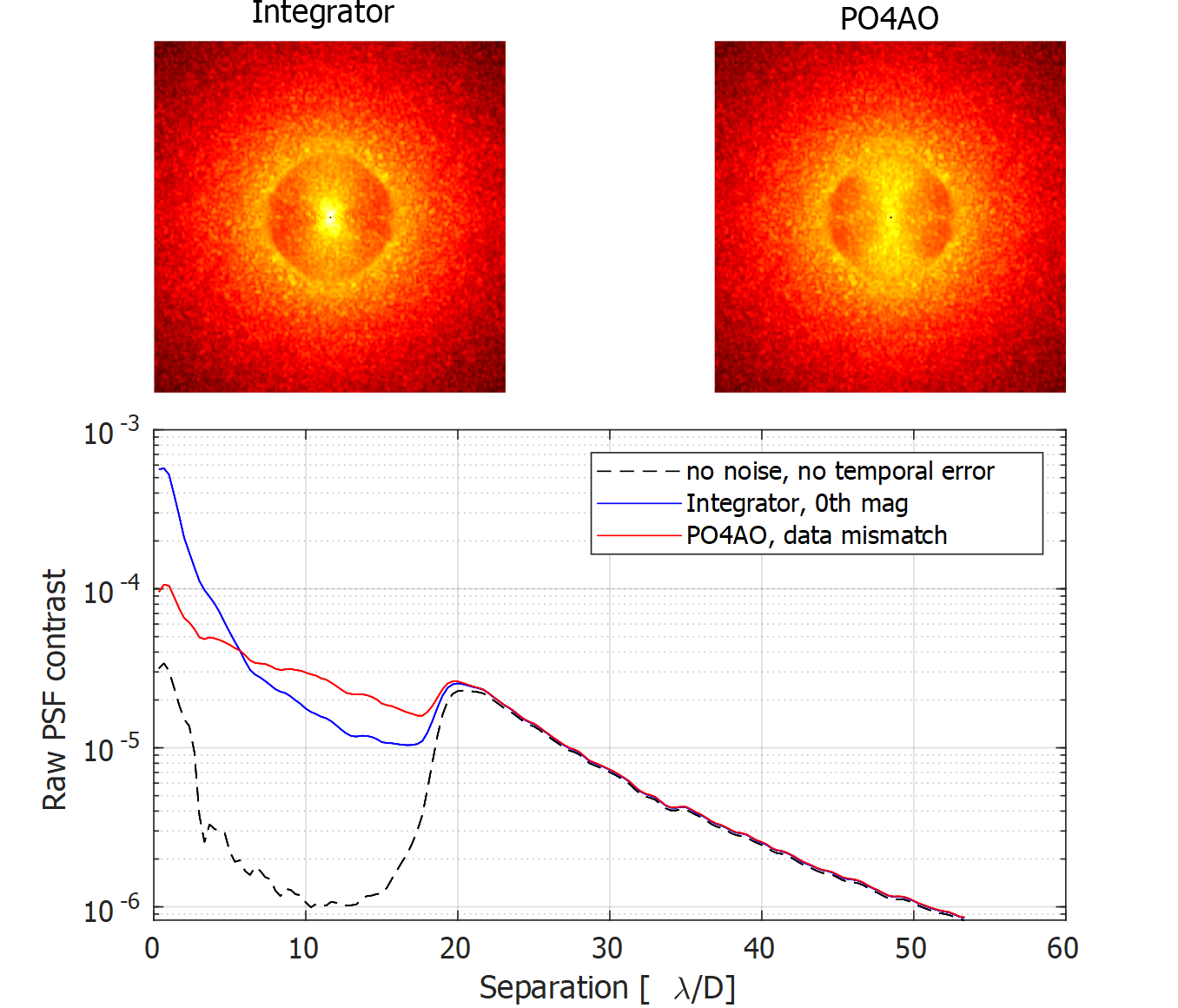}}
      
\caption[Raw PSF contrast]
        {Raw PSF contrast in VLT-scale telescope experiments. Upper images: Raw PSF contrast. Lower plot: The radial averages over the image. The blue lines are for the integrator and red for the PO4AO. The raw PSF contrast was computed during the 1000 frames of the experiment.}
    \label{fig:VLT contrast}
\end{figure*}

\begin{figure}
\centering
\includegraphics[trim={1.cm 0cm 2cm 0cm}, clip, width=0.48\textwidth]{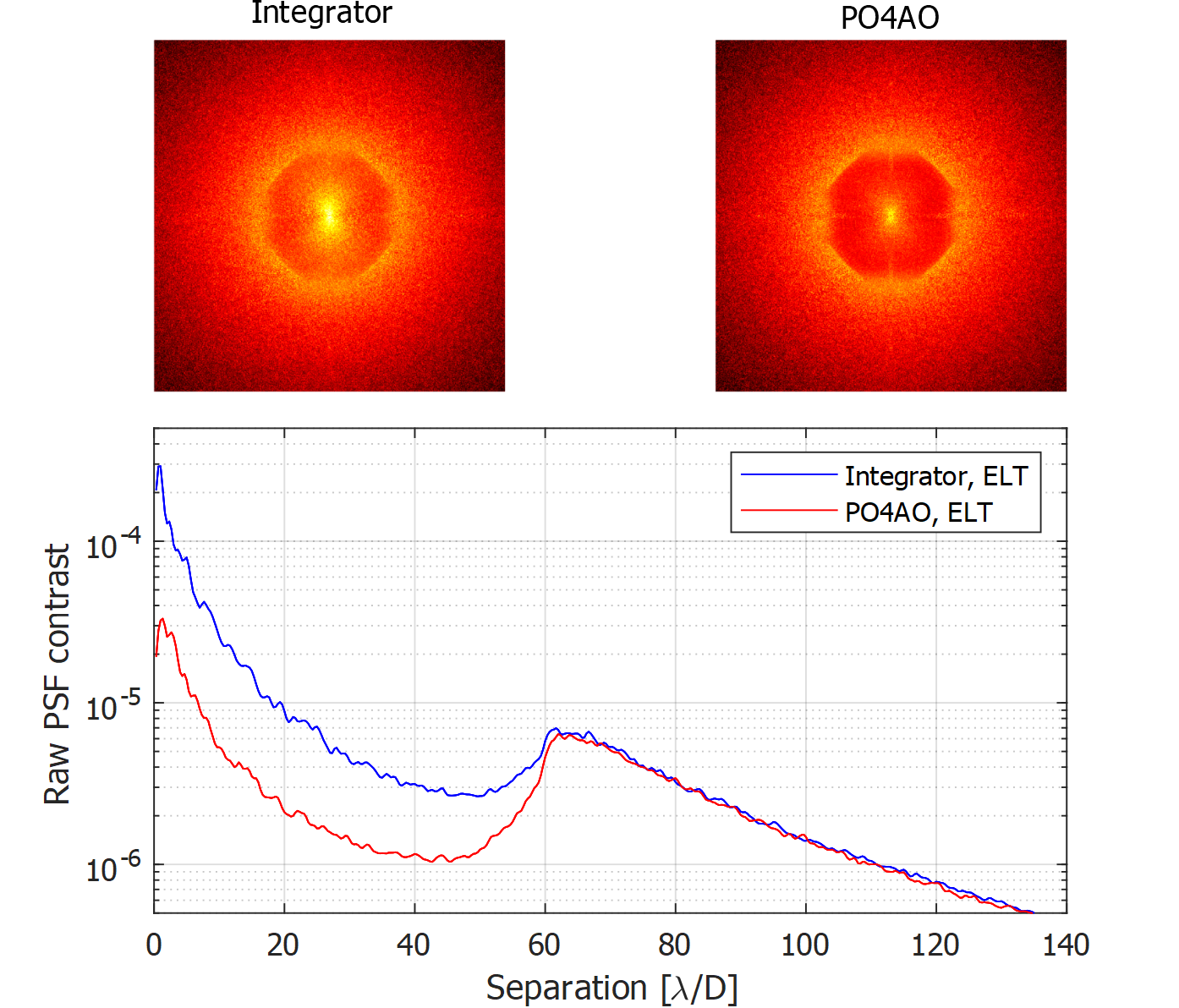}%
\caption{Raw PSF contrast in the ELT-scale experiment.} 
\label{fig:pcs_contrast}
\end{figure}

\section{MagAO-X}
In addition to running the numerical simulations presented in the previous section, we also implemented PO4AO on the MagAO-X instrument. MagAO-X is an experimental coronagraphic extreme adaptive optics system that uses woofer-tweeter architecture (ALPAO-97 DM as the woofer and  Boston Micromachines 2K as the tweeter). We use a point source in the f/11 input focus to illuminate the DMs, Pyramid WFS, and scientific camera. Further, we place a classical Lyot coronagraph with a 2.5 $\lambda/D$ Lyot mask radius in front of the science camera. We set PWFS's modulation ratio to $3 \lambda/D$, and the brightness of the guide star is adjusted to match the flux per frame which a 0th magnitude star would provide in 1 ms, i.e., for a system running at 1 kHz. We used a similar test setup as \cite{haffert2021data1} and ran our experiment by only controlling the woofer DM and injecting disturbances by running simulated phase screens across it. The phase screens were simulated as single-layer frozen flow turbulence with $r_0$ of 16 cm at $500$nm. We experimented with three different single-layer wind profiles: 5 m/s, 15 m/s, and 30 m/s, where the wind speeds correspond to a 1 Khz framerate again.

The PO4AO is implemented with PyTorch and utilizes the Python interface of the MagAO-X RTC to pass data from CPU to GPU memory, do the PO4AO calculations on the GPU, and transfer them back. The data transfer takes time and limits the achievable framerate in this setup to 100 Hz. RTC software that would run entirely on GPUs would not suffer from this limitation.

\subsection{The integrator}
To retrieve the interactions matrix, we used the standard calibrations process described in Section 3. From the interactions matrix, we derived the reconstruction matrix by Tikhonov regularization given by,

\begin{equation}
C = (D^\top D +\alpha I  )^{-1}D^\top,  
\end{equation}
where $\alpha$ is tuned manually. We also tuned the integrator gain manually for each wind profile. 

\subsection{PO4AO}
The structure of the MagAO-X experiment is similar to our numerical simulations. First, we trained the PO4AO for 50 episodes ($25000$ frames) and then ran for an additional 5000 frames to compare the post-coronagraphic PSFs. We also use the $10$ episode warmup with noisy integrator and the same NN architectures. Given the low number of actuators and the high-order PWFS, we set the number of past telemetry data (k and m) to 10, and instead of filtering ~20\% of the K-L modes, for maximum performance, we only filter the piston mode in the policy output (see Figure \ref{fig:models}).

\subsection{Results}
\begin{figure*}[!t]
    \begin{minipage}[b][15cm]{0.5\linewidth}
        \includegraphics[width=.9\linewidth]{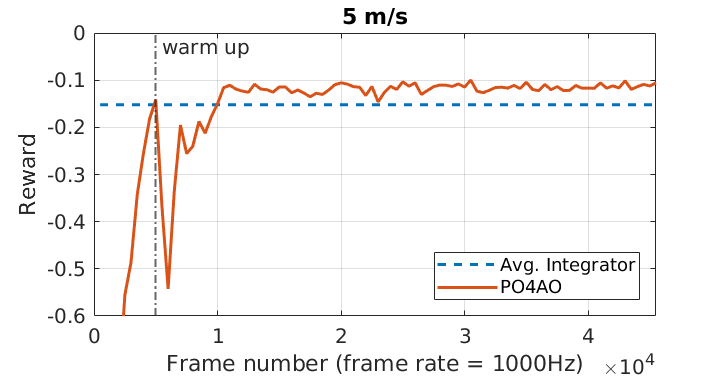}
        \vfill
        \includegraphics[width=.9\linewidth]{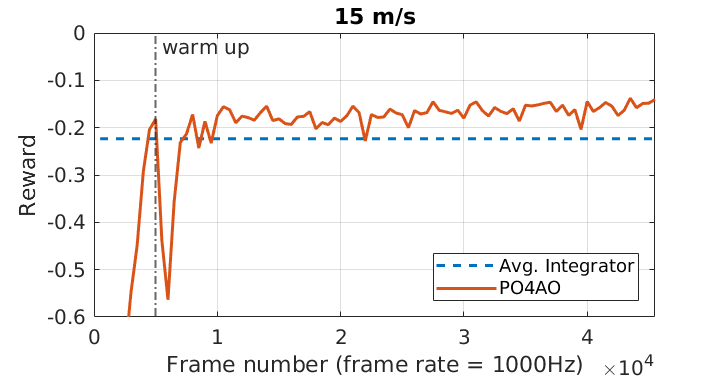}
        \vfill
        \subfloat[Training curves for PO4AO in the lab setup. The red lines are for PO4AO performance and dashed blue line represents the average integrator performance over an episode. The dashed gray vertical line is where the policy is switch from noisy integrator to PO4AO. For all different wind conditions the PO4AO passes the integrator performance after 10k frames of data. \label{fig:train_mag}]{\includegraphics[width=.9\linewidth]{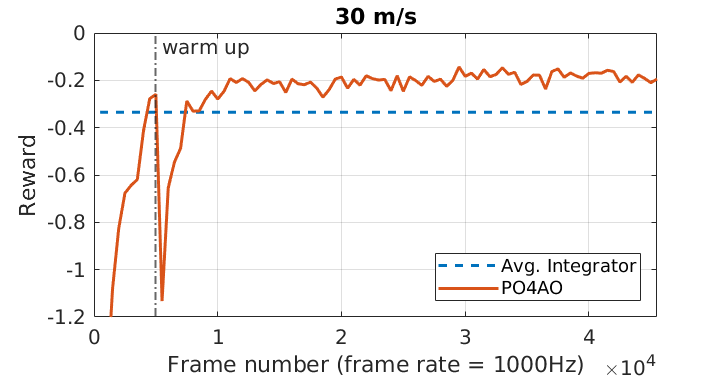}}
    \end{minipage}%
    \hfill
    \begin{minipage}[b][15cm]{0.5\linewidth}
        \begin{center}
            \subfloat[MagAO-X post-coronagraphic PSFs of the methods. Left is for the Integrator and right for the PO4AO. The PSFs are limited by NCPA and, in order to validate the method, we examined the temporal variance of the PSFs (see Figure \ref{fig:magaox}) \label{fig:magaox_psf}]{\includegraphics[trim={0cm 0cm 0cm 0cm}, clip,width=0.9\textwidth]{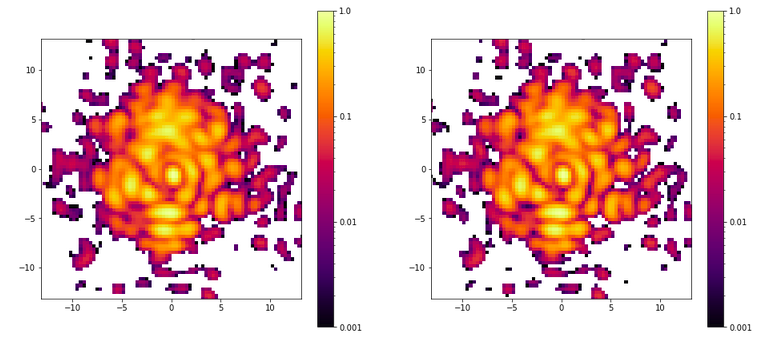}}
        \end{center}
        \vfill
        \subfloat[Temporal variance of MagAO-X post-coronagraphic PSFs. Upper images: Temporal speckle variance at image plane for both control methods (left: integrator, right: PO4AO). Lower image: radial average over the images. The blue line is for the integrator and the red line for the PO4AO. The gray vertical line represents the inner working angle of the coronagraph (radius $2.5 \lambda/D$). \label{fig:magaox}]{\includegraphics[trim={1.cm 0cm 2cm 0cm}, clip, width=0.96\textwidth]{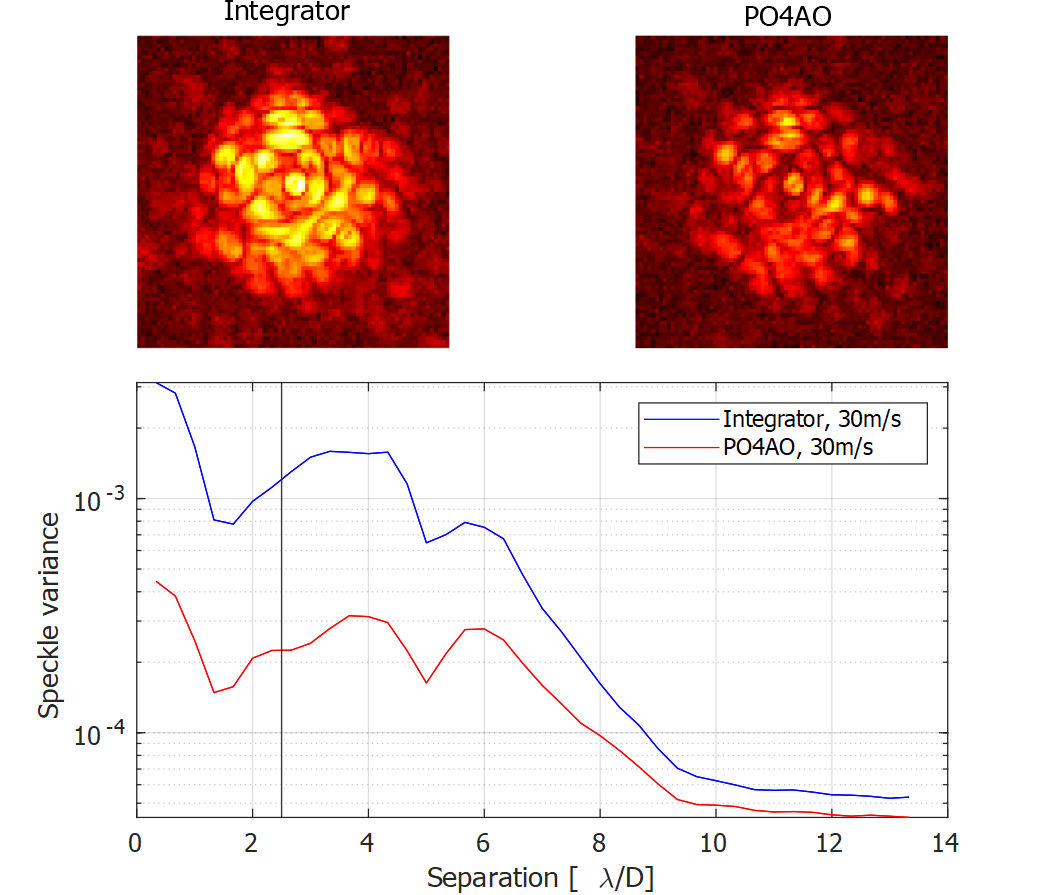}}
    \end{minipage}%
    \caption{MagAO-X experiment results.} 
    \label{fig:magaox_total}
\end{figure*}

We compare the performance of the PO4AO to the integrator in two ways: by looking at the training curves (see Figure \ref{fig:train_mag}) and by comparing the post-coronagraphic speckle variance (see Figure \ref{fig:magaox}). The PO4AO achieves better performance in all wind conditions than the integrator after 10k (10s in theoretical real-time) data frames. The reward is proportional to the mean RMS of the reconstructed wavefront. We further examine the performance by comparing the post-coronagraphic images' with the 30 $m/s$ wind profile; see Figures \ref{fig:magaox_psf} and \ref{fig:magaox}. The residual intensities of the images' (see Figure \ref{fig:magaox_psf}) are limited by NCPA. Therefore, instead of comparing the raw PSF contrast, we compare the temporal speckle variance of the method (see Figure \ref{fig:magaox}). We see a factor of $3-7$ improvement in the speckle variance at  $2.4-6 \lambda/D$. Given the inner working angle of the coronagraph and DM's control radius, that is where we would also expect to see the improvement. Further, these results are in line with the results from numeric simulations.

\section{Discussion}
In conclusion, reinforcement learning is a promising approach for AO control that could be implemented in on-sky systems with already existing hardware. The algorithm we propose requires only a small amount of training data and maintains an acceptable performance even when the training conditions differ heavily from test time. Further, it has a high inference speed, capable of scaling to high-order instruments with up to 10k actuators. Thanks to the use of relatively shallow convolutional NN, the inference time is just 300 $\mu s$ with a modern laptop GPU. The inference time is also similar for an ELT scale system with more than 10k actuators and for a VLT scale system with 'just' 1400 actuators.  

The method was tested in numerical simulations and a lab setup and provided significantly improved post-coronagraphic contrast for both cases compared to the integrator. It is entirely data-driven, and in addition to predictive control, it can cope with modeling errors such as the optical gain effect and highly non-linear wavefront sensing. Due to the constantly self-calibrating nature of the algorithm it could turn AO control into a turnkey operation, where the algorithm maintains itself entirely automatically.

We showed that our method is robust to heavy data mismatch, but the performance is reduced for a short time while PO4AO is adapting to the evolution of external conditions. These abrupt changes in wind conditions will rarely occur in the real atmosphere. Therefore, future work should also address maintaining the best possible performance under reasonably varying turbulence. The model learns on a scale of several seconds and can presumably adapt to changing atmospheric conditions at the same time scale. However, more research on the trade-off between model complexity and training speed is still needed. For example, a deeper NN model could generalize better to unseen conditions, while shallower NN models could learn new unseen conditions faster. Currently, the CNN model architectures themselves are not thoroughly optimized, and an exciting research topic would be to find the optimal CNN design to capture the AO control system dynamics for model-based RL. For example, a U-net type CNN architectures \citep{ronneberger2015u} and mixed-scale dense CNNs \citep{pelt2018mixed} have shown excellent performance on imaging-related applications. On the other hand, we could utilize similar NN structures that have shown excellent performance in pure predictive control \citep{swanson2018wavefront, swanson2021closed}. Such a study should consider a variety of different, preferably realistically changing atmospheric conditions and misalignments as well as prerecorded on-sky data.

As a caveat, the algorithm, like most deep RL methods, is somewhat sensitive to the choice of hyperparameters (e.g., number of layers in neural networks, learning rates, etc.). Moreover, control via deep learning is hard to analyze, and no stability bounds can be established. 

Further, development is needed to move from the laboratory to the sky. The method currently runs on a Python interface that has to pass data via the CPU on MagAO-X. To increase the speed of the implementation and the maximum framerates, we must switch to a lower-level implementation that runs both the real-time pipeline and the PO4AO control on the GPU using the same memory banks. In addition, the training procedure needs to run in parallel with the inference, which should be straightforward to implement. 

To summarize, this work presents a significant step forward for XAO control with RL. It will allow us to increase the S/N, detect fainter exoplanets, and reduce the time it takes to observe them on ground-based telescopes. As astronomical telescopes become larger and larger, the choice of the AO control method becomes critically important, and data-driven solutions are a promising direction in this line of work. Deep learning and RL methods are transforming many fields, such as protein folding, inverse problems, and robotics, and there is potential for the same to happen for direct exoplanet imaging.

\bibliographystyle{aa}
\bibliography{aanda}

\end{document}